\documentclass[a4paper,12pt]{article}
\usepackage{amsmath,natbib,comment,graphicx,bm,color}
\usepackage[driverfallback=dvipdfmx]{hyperref}
\usepackage{subfigure}

\newtheorem{Remark}{Remark}[section]

\def\vtheta{\bm{\theta}}

\def\vx{\bm{x}}

\def\vpsi{\bm{\psi}}
\def\vmu{\bm{\mu}}
\def\vs{\bm{s}}

\def\gmethod{$\gamma$-lasso}
\def\ee{\varepsilon}
\def\tlasso{$t$lasso}

\begin{document}

\begin{center}
\textbf{\Large  Robust sparse Gaussian graphical modeling}
\end{center}
\begin{center}
\large {Kei Hirose$^1$, Hironori Fujisawa$^2$ and Jun Sese$^3$}
\end{center}

\begin{center}
{\it {\small
$^1$ Institute of Mathematics for Industry, Kyushu University,\\
744 Motooka, Nishi-ku, Fukuoka 819-0395, Japan. \\

\vspace{1.2mm}

$^2$ The Institute of Statistical Mathematics,\\
10-3 Midori-cho, Tachikawa, Tokyo 190-8562, Japan.\\

\vspace{1.2mm}

$^3$ National Institute of Advanced Industrial Science and Technology, \\
2-4-7 Aomi, Koto-ku, Tokyo 135-0064, Japan

}}
{\it {\small E-mail: hirose@imi.kyushu-u.ac.jp, fujisawa@ism.ac.jp, sese.jun@aist.go.jp
}}
\end{center}
\begin{abstract}
Gaussian graphical modeling has been widely used to explore various network structures, such as gene regulatory networks and social networks.  We often use a penalized maximum likelihood approach with the $L_1$ penalty for learning a high-dimensional graphical model.  However, the penalized maximum likelihood procedure is sensitive to outliers.  To overcome this problem, we introduce a robust estimation procedure based on the $\gamma$-divergence.  The proposed method has a redescending property, which is known as a desirable property in robust statistics.  The parameter estimation procedure is constructed using the Majorize-Minimization algorithm, which guarantees that the objective function monotonically decreases at each iteration.  Extensive simulation studies showed that our procedure performed much better than the existing methods, in particular, when the contamination ratio was large. Two real data analyses were carried out to illustrate the usefulness of our proposed procedure.  
\end{abstract}
\noindent%

\section{Introduction}
Gaussian graphical modeling has been widely used to investigate the conditional independence between two variables given other variables.  Under a Gaussian assumption, the  conditional independence between two variables corresponds to the zero entry of inverse covariance matrix \citep{edwards2000introduction}.  A sparse estimation of the inverse covariance matrix, i.e., a method in which some of the elements of the inverse covariance matrix are shrunk to exactly zero, is often used to obtain the conditional independence graph.

 In many applications, the number of variables is much larger than the number of observations.  An example is the analysis of microarray gene expression data, in which we discover the relation between pairs of genes.   In such a case,  the maximum likelihood estimate of the inverse covariance matrix does not exist.  To overcome this problem, there has been a great deal of interest on the $L_1$ regularization, such as the lasso \citep{Tibshirani:1996}, for estimating the sparse inverse covariance matrix.  \citet{meinshausen2006high} proposed fitting the lasso regression to each variable, in which one variable is a response and the other variables are predictors.  Their method does not guarantee that the non-zero pattern of the inverse covariance matrix is symmetric.  \citet{peng2009partial} introduced a joint regression, which is also based on the lasso regression and ensures the symmetry of the estimated inverse covariance matrix.  \citet{yuan2007model} considered the problem of maximizing the penalized log-likelihood function via the lasso (hereafter referred to as the {\it graphical lasso}).  Among these methods, the graphical lasso has been becoming popular because of its computational efficiency (e.g., \citealp{,friedman2008sparse,witten2011new,hsieh2011sparse}) and desirable statistical properties in high-dimensional settings \citep{rothman2008sparse,raskutti2008model}.

In practical situations, however, outliers are often observed or the distribution is heavy-tailed \citep{finegold2011robust,fritsch2012detecting}.  In such cases, the conventional estimation procedure may produce an inappropriate graph structure.  To overcome this problem, a few researchers proposed the robust estimation procedures with the $L_1$ penalization.  \citet{liu2009nonparanormal} proposed the {\it nonparanormal}, in which a truncated marginal empirical distribution was adopted to remove outliers and a semiparametric Gaussian copula was used to treat a conditional independence structure. The model parameter was estimated by a standard algorithm of the graphical lasso, such as the blockwise coordinate descent algorithm \citep{friedman2008sparse}.  \citet{finegold2011robust} introduced the {\it \tlasso}, in which the underlying distribution was assumed to be the multivariate $t$-distribution with a heavy tail. The model parameters were estimated by the EM algorithm. \citet{vinciotti2013robust} compared performances of various robust estimation procedures, including the nonparanormal and the $t$lasso, and they concluded the nonparanormal performed well in various situations.   \citet{sun2012robust} considered a modified likelihood approach based on the density power divergence  \citep{basu1998robust}  (hereafter referred to as the {\it dp-lasso}). 
The model parameter was estimated by the coordinate descent algorithm with a quadratic approximation \citep{tseng2009coordinate}.  

However, the above procedures have some drawbacks.  
The nonparanormal approach removes observations on both sides at the ratio $2\delta$, i.e., observations that have extremely large positive and negative values are removed at the same ratio $\delta$.  The truncation parameter $\delta$ corresponds to the contamination ratio and must be selected beforehand. \cite{liu2009nonparanormal} selected $\delta$ such that it achieved a desired rate of convergence of the estimator. Nevertheless, the selected truncation parameter tends to be too small when the contamination ratio is large, because $\delta \rightarrow 0$ as $n \rightarrow \infty$, where $n$ is the number of the observations. In addition, 
the outliers may not exist on both sides at the same ratio $\delta$.  In fact, the outliers of yeast gene expression data described in Section 6.1 have only large negative values.   The \tlasso\ has the same drawback, because the heavy tail distribution implies that outliers are assumed on both sides.  Furthermore, the \tlasso\ tends to yield a large variance of the estimator, because a heavier tail distribution often produces a smaller Fisher information.   The dp-lasso approach has four tuning parameters to be determined, and is often unstable in our experience.  In our simulation study, we observed that the above three estimation procedures performed poorly when the contamination ratio was large and the outliers were present on one side, and the estimators had large mean squared errors even when the number of observations was sufficiently large. 

To handle the issues above, we propose the {\it \gmethod}, which is a robust sparse estimation procedure of the inverse covariance matrix based on the $\gamma$-divergence \citep{fujisawa2008robust,cichocki2010families}.   The \gmethod\ regards an observation whose likelihood value is small as an outlier, unlike the nonparanormal. As a result, the \gmethod\ can appropriately treat the outliers even when they exist on only one side. In addition, we do not need to know the contamination ratio in advance.  The $\gamma$-lasso tends to yield a much smaller variance of the estimator than the \tlasso, because the underlying distribution is assumed to be Gaussian.  The parameter estimation algorithm is proposed using the Majorize-Minimization algorithm (MM algorithm, \citealp{hunter2004tutorial}), which guarantees that the objective function monotonically decreases at each iteration.  The proposed algorithm does not have any tuning parameters to be determined. As a result, the parameter estimation is more stable than the dp-lasso.  In addition, the \gmethod\ has a redescending property, which is known as a desirable property in robust statistics, so that the bias of the estimator is expected to be sufficiently small when an outlier takes a large value \citep{maronna2006robust}.  We conducted extensive Monte Carlo simulations to investigate the performance of the proposed procedure.  The result showed that our procedure performed  better than existing methods in most cases.    The proposed procedure is available for use in the {\tt R} package {\tt rsggm}\footnote{Available at \url{http://cran.r-project.org/web/packages/rsggm.}}.

The organization of this paper is given as follows. In Section 2, we introduce a robust estimation of the sparse inverse covariance matrix via the $\gamma$-divergence.   Section 3 provides a parameter estimation procedure via the MM algorithm.  In Section 4, we compare the proposed procedure with several existing methods.  Section 5 investigates the effectiveness of our proposed procedure via Monte Calro simulations.  Section 6 describes two real data analyses of the gene expression data.  Concluding remarks are given in Section 7.  Some technical proofs are collected in Appendices.  

\section{Robust and sparse estimation of the inverse covariance matrix}
\subsection{Gaussian graphical model}
Let $\bm{X}=(X_1,\dots,X_p)^T$ be the $p$-dimensional multivariate-normally distributed random variable with mean vector $\bm{\mu} = (\mu_1,\dots,\mu_p)^T$ and covariance matrix $\bm{\Sigma} = (\sigma_{ij})$. Let the inverse covariance matrix of $\bm{\Sigma}$ be denoted by $\bm{\Omega}=(\omega_{ij})$. It is well-known that each variable is written as $X_i = \sum_{j \ne i} \beta_{ij}X_j + \delta_i$, where $\beta_{ij} = -\frac{\omega_{ij}}{\omega_{ii}}$ and $\delta_i \sim N(0,1/\omega_{ii})$, and then the zero/non-zero element of the inverse covariance matrix corresponds to the conditional independence/dependence given other variables. The sparsity pattern of the inverse covariance matrix corresponds to the graph structure: there is an edge between $i$th and $j$th vertices if and only if $\omega_{ij} \ne 0$.  We estimate the inverse covariance matrix by a sparse matrix to obtain a sparse graphical model.

\subsection{Sparse estimation of the Gaussian graphical model}
Suppose that we have a random sample of $n$ observations $\bm{x}_1,\cdots,\bm{x}_n$ from the $p$-dimensional normal population $N(\bm{\mu} ,\bm{\Sigma} ) $.  To estimate the sparse inverse covariance matrix, \citet{yuan2007model} proposed minimizing the following penalized negative log-likelihood function:
\begin{eqnarray*}
\ell_{\lambda}(\bm{\theta}) = \ell(\bm{\theta}) + \frac{\lambda}{2} \| \bm{\Omega} - {\rm diag}(\bm{\Omega}) \|_1, \label{解きたい問題}
\end{eqnarray*}
where $\ell(\bm{\theta})$ is a negative log-likelihood function given by $\ell(\bm{\theta}) = - \sum_{i=1}^n \log f(\bm{x}_i;\bm{\theta})$, and $\lambda\ge 0$ is a tuning parameter which controls the balance between sparsity of parameters and goodness of fit to the data.  
Here $f(\bm{x};\bm{\theta})$ is a density function of the multivariate normal distribution
\begin{equation*}
	f(\bm{x};\bm{\theta}) = (2\pi)^{-p/2}|\bm{\Omega}|^{1/2}\exp\left( -\frac{1}{2}(\bm{x} - \bm{\mu})^T\bm{\Omega}(\bm{x} - \bm{\mu}) \right),
\end{equation*}
and $\bm{\theta}$ is a model parameter expressed as $\bm{\theta} = (\bm{\mu}^T,{\rm vech}(\bm{\Omega})^T)^T$.  

For any $\bm{\Omega}$, the penalized negative log-likelihood function $\ell_{\lambda}(\bm{\theta})$ is minimized when the mean vector $\bm{\mu}$ is the sample mean $\hat{\bm{\mu}} = \frac{1}{n} \sum_{i=1}^n \bm{x}_i $. The inverse covariance matrix $\bm{\Omega}$ is then estimated by the following minimization problem:
\begin{eqnarray}
\min_{\bm{\Omega}} \left\{ - \log |\bm{\Omega} | +{\rm tr} (\bm{\Omega}\bm{S}) + \lambda \| \bm{\Omega} - {\rm diag}(\bm{\Omega}) \|_1 \right\}, \label{解きたい問題_adj}
\end{eqnarray}
where $\bm{S}=(s_{ij})$ is the sample covariance matrix. The problem of (\ref{解きたい問題_adj}) is referred to as the {\it graphical lasso} \citep{witten2011new}.  Several researchers have proposed efficient algorithms for solving the problem of (\ref{解きたい問題_adj}), such as the blockwise coordinate descent approach \citep{friedman2008sparse}, the quadratic approximation \citep{hsieh2011sparse}, and the Alternating Direction Method of Multipliers (ADMM; \citealp{boyd2011distributed}).

\subsection{Robust estimation via the $\gamma$-divergence}
In practical situations, the estimate obtained by (\ref{解きたい問題_adj}) is sensitive to outliers.  To obtain a robust estimate, instead of the negative log-likelihood function $\ell(\bm{\theta})$, we use the {\it negative $\gamma$-likelihood function} \citep{fujisawa2008robust,cichocki2010families}, given by
\begin{eqnarray}
\ell_{\gamma}(\bm{\theta}) = \mathop{\underline{-\frac{1}{\gamma} \log \left\{ \frac{1}{n}\sum_{i=1}^n f(\bm{x}_i;\bm{\theta})^{\gamma}  \right\}}}_{\ell_1(\bm{\theta})} 
+ \mathop{\underline{\frac{1}{1+\gamma} \log \int f(\bm{x};\bm{\theta})^{1+\gamma}d\bm{x}}}_{\ell_2(\bm{\theta})},
\label{ガンマダイバージェンス}
\end{eqnarray}
where $\gamma \ge 0$ is a tuning parameter which controls the balance between efficiency and robustness.  Note that $\gamma \rightarrow +0$ corresponds to the negative log-likelihood function.  The first term $\ell_1(\bm{\theta})$ can lead to a robust estimation, and the second term $\ell_2(\bm{\theta})$ makes the bias of the estimate sufficiently small.

We provide an intuitive explanation about how the negative $\gamma$-likelihood function can lead to the robust estimation.  Suppose that $\bm{x}_1$ is an outlier.  The likelihood $f(\bm{x}_1;\bm{\theta})$ is expected to be sufficiently small.  Minimizing the negative log-likelihood function $\ell(\bm{\theta}) = - \sum_{i=1}^n \log f(\bm{x}_i;\bm{\theta})$ cannot make $f(\bm{x}_1;\bm{\theta})$ extremely small, because $\ell(\bm{\theta}) \rightarrow \infty$ as $f(\bm{x}_1;\bm{\theta}) \rightarrow 0$.  On the other hand, with the negative $\gamma$-likelihood function $\ell_{\gamma}(\bm{\theta})$, the likelihood term $f(\bm{x}_1;\bm{\theta})$ is naturally ignored, because we can easily obtain the following approximation: 
\begin{eqnarray*}
\arg\min_{\bm{\theta}} \ell_\gamma(\bm{\theta}) 
  \approx \arg\min_{\bm{\theta}} \left[ - \frac{1}{\gamma} \log \left\{ \frac{1}{n-1}\sum_{i=2}^n f(\bm{x}_i;\bm{\theta})^{\gamma}\right\} + \frac{1}{1+\gamma} \log \int f(\bm{x};\bm{\theta})^{1+\gamma}d\bm{x} \right].
\end{eqnarray*}

The robust and sparse estimate is proposed by
\begin{eqnarray*}
\hat{\bm{\theta}}= \arg\min_{\bm{\theta}} \ell_\gamma(\bm{\theta}),
\end{eqnarray*}
where
\begin{eqnarray}
\ell_{\gamma,\lambda}(\bm{\theta}) = \ell_{\gamma}(\bm{\theta}) + \frac{\lambda}{2}  \| \bm{\Omega} - {\rm diag}(\bm{\Omega}) \|_1.  \label{罰則付きガンマダイバージェンス}
\end{eqnarray}
We call $\ell_{\gamma,\lambda}(\bm{\theta})$ the {\it penalized negative $\gamma$-likelihood function}, and the minimization problem of (\ref{罰則付きガンマダイバージェンス}) the {\it $\gamma$-lasso}.

 \subsection{Illustrative Example}
\begin{figure}[!htbp]
\centering
\subfigure[graphical lasso applied to the uncontaminated data]{\includegraphics[width=5cm]{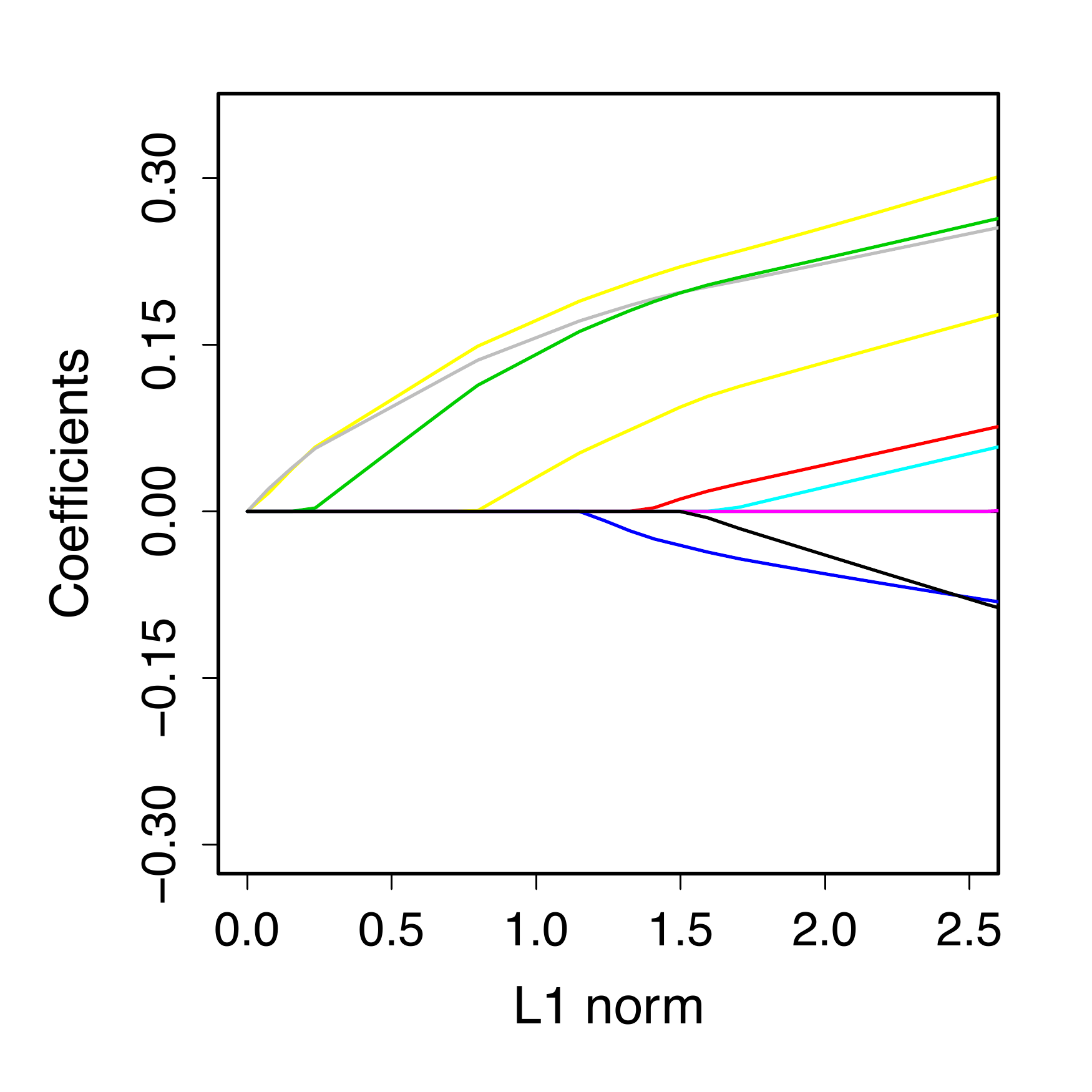}
 \label{fig:sim_solpath_lasso_nooutlier}}
\subfigure[graphical lasso]{\includegraphics[width=5cm]{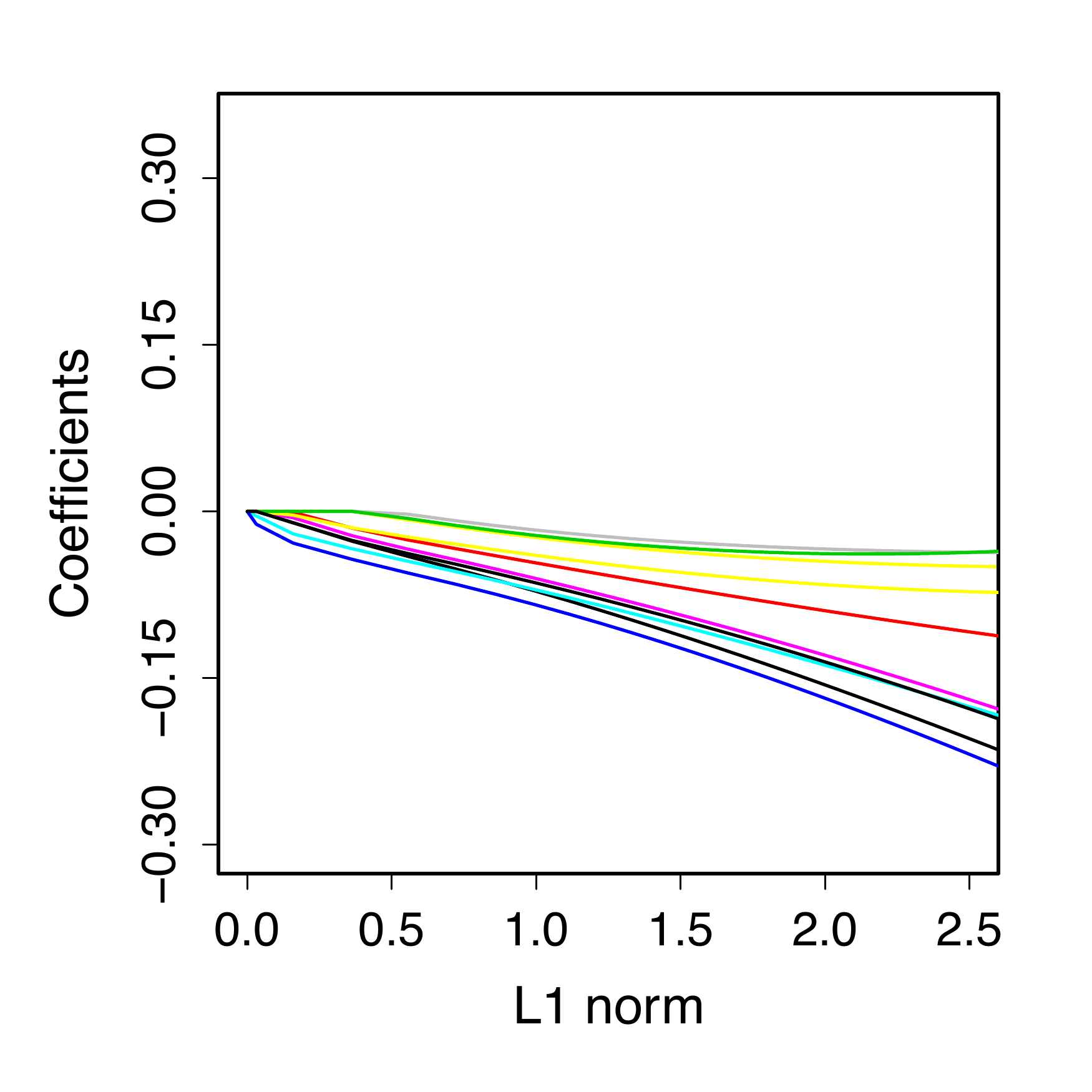}
 \label{fig:sim_solpath_lasso}}
\subfigure[$\gamma$-lasso]{\includegraphics[width=5cm]{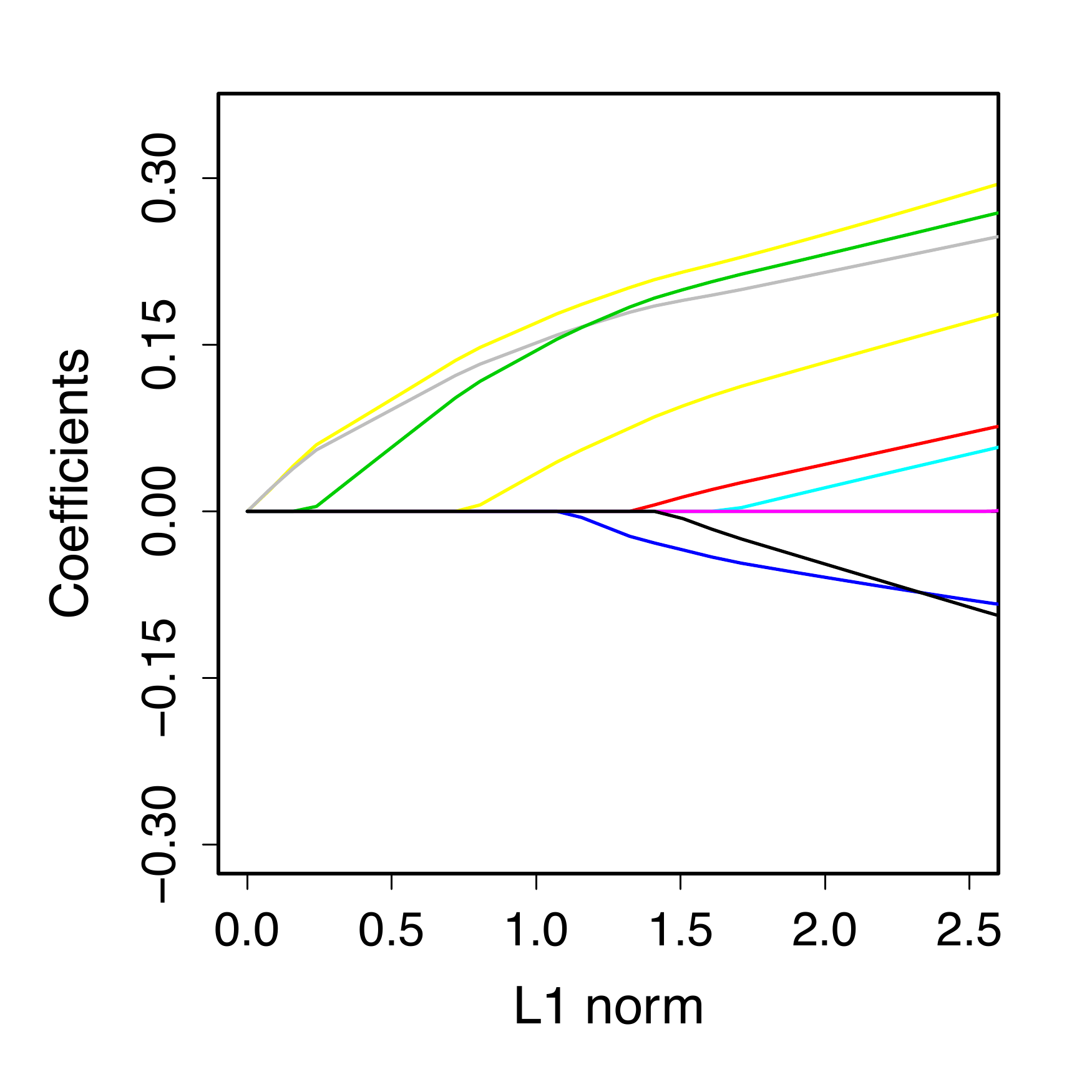}
 \label{fig:sim_solpath_gamma-lasso}}
\caption{Solution paths (estimates of $\omega_{ij}$ ($i,j=1,\dots, p$, $i<j$) as a function of $\sum_{i < j}|\hat{\omega}_{ij}|$) for the dataset generated from a mixture distribution. (a) The solution path of the graphical lasso applied to the uncontaminated data in which the outliers were removed from the original contaminated data. (b)-(c) The solution paths of graphical lasso and \gmethod\ applied to the original contaminated data.  }
 \label{fig:solutionpath_simulation}
\end{figure}

We provide solution paths (estimates of $\omega_{ij}$ ($i,j=1,\dots, p$, $i<j$) as a function of $\sum_{i < j}|\hat{\omega}_{ij}|$) of the $\gamma$-lasso and the ordinary graphical lasso when outliers exist.  We generated $n=200$ observations from a mixture distribution $0.9N(\bm{0},\bm{\Omega}^{-1}) + 0.1N(\bm{5},\bm{I})$, where 
$$
\bm{\Omega} = 
\begin{pmatrix}
 1.0 & 0.3 & 0.3 & 0.0 & 0.0 \\ 
 0.3 & 1.0 & 0.0 & 0.0 & 0.3 \\ 
 0.3 & 0.0 & 1.0 & 0.3 & 0.0 \\ 
 0.0 & 0.0 & 0.0 & 1.0 & 0.0 \\ 
 0.0 & 0.3 & 0.0 & 0.0 & 1.0 \\ 
\end{pmatrix},
$$
 $\bm{I}$ is an identity matrix, and $\bm{5}$ is a $5$-dimensional vector whose elements are 5.   In this case, $N(\bm{0},\bm{\Omega}^{-1})$ is our target model and $N(\bm{5},\bm{I})$ corresponds to the contamination.  Figure~\ref{fig:solutionpath_simulation}(a) shows the solution path of the graphical lasso applied to the uncontaminated data in which the outliers were removed from the original contaminated data.  Figures~\ref{fig:solutionpath_simulation}(b) and (c) depict the solution paths of graphical lasso and \gmethod, respectively, applied to the original contaminated data.  For \gmethod, we chose $\gamma=0.1$.

 Clearly, the solution path of the graphical lasso in Figure \ref{fig:solutionpath_simulation}(b) was completely different from Figure~\ref{fig:solutionpath_simulation}(a), which implies that the graphical lasso was highly sensitive to the outliers.  However, the solution path of the \gmethod\ in Figure \ref{fig:solutionpath_simulation}(c) was almost the same as Figure~\ref{fig:solutionpath_simulation}(a), so that our method was robust against the outliers. 

\subsection{Redescending Property}
Suppose that the estimating equation is given by $\sum_{i=1}^n \vpsi(\vx_i;\vtheta)=\bm{0}$.  The estimating equation is said to have a redescending property if $\lim_{\|\vx\| \rightarrow \infty} \bm{\psi}(\vx;\vtheta^*) = \bm{0}$, where $\vtheta^*$ is a true parameter of $\vtheta$.  The redescending property on M-estimation (see, e.g.,  \citealp{maronna2006robust}) is  known as a desirable property in robust statistics, because the bias of the estimator is expected to be sufficiently small when an outlier takes a large value \citep{maronna2006robust}. 

The $\gamma$-lasso minimizes the loss function in (\ref{罰則付きガンマダイバージェンス}), so that the estimating equation is expressed as
\begin{eqnarray}
-\dfrac{\sum_{i=1}^n f(\vx_i;\vtheta)^\gamma \vs (\vx_i;\vtheta)}{\sum_{j=1}^n f(\vx_j;\vtheta)^\gamma } + \frac{\partial}{\partial\vtheta} \ell_2(\vtheta) + \frac{\lambda}{2} \bm{u} = \bm{0}, \label{estimating equation}
\end{eqnarray}
where
$$ \vs(\vx;\vtheta) = \frac{\partial \log f(\vx;\vtheta)}{ \partial \vtheta}, \quad \bm{u}=(u_1,\dots,u_{p(p+2)})^T, \quad u_j \in [-1,1].
$$
Therefore, the kernel function $\vpsi(\vx;\vtheta)$ is given by\begin{eqnarray}
\vpsi(\vx;\vtheta) = f(\vx;\vtheta)^\gamma \vs (\vx;\vtheta) - f(\vx;\vtheta)^\gamma \frac{\partial}{\partial\vtheta} \ell_2(\vtheta) - f(\vx;\vtheta)^\gamma \frac{\lambda}{2} \bm{u}. \label{kernel function}
\end{eqnarray}
Because $f(\vx;\vtheta)$ is the density function of the Gaussian distribution, we have 
\begin{eqnarray*}
f(\vx;\vtheta)^{\gamma} \frac{\partial \log f(\vx;\vtheta)}{ \partial \vmu} &=& (2\pi)^{-p\gamma/2} |\bm{\Omega}|^{\gamma/2} \exp\left\{ -\frac{\gamma}{2} (\vx-\vmu)^T \bm{\Omega} (\vx-\vmu) \right\} \\
 & & \hspace{10mm} \times \left[\bm{\Omega} (\vx-\vmu) \right] \ \rightarrow \bm{0} 
 \hspace{33mm}\mbox{(as }\|\bm{x}\| \rightarrow \infty), \\
f(\vx;\vtheta)^{\gamma} \frac{\partial \log f(\vx;\vtheta)}{ \partial \omega_{jk}} &=& (2\pi)^{-p\gamma/2} |\bm{\Omega}|^{\gamma/2} \exp\left\{ -\frac{\gamma}{2} (\vx-\vmu)^T \bm{\Omega} (\vx-\vmu) \right\} \\
 & & \hspace*{10mm} \times \left[ \frac{1}{2} \frac{\partial}{\partial \omega_{jk}} \log |\bm{\Omega}| - \frac{1}{2} (\vx-\vmu)^T \frac{\partial\bm{\Omega}}{\partial \omega_{jk}} (\vx-\vmu) \right] \ \rightarrow \bm{0} \\ 
  && \hspace{80mm}\mbox{(as }\|\bm{x}\| \rightarrow \infty).
\end{eqnarray*}
Therefore, the first term of the right side in (\ref{kernel function}) approaches 0 as $\|\vx\| \rightarrow \infty$.  The elements of $\bm{u}$ are bounded, so that the second and third terms of the right side in (\ref{kernel function}) also approach 0 as $\|\vx\| \rightarrow \infty$.  As a result, the estimating equation has a redescending property. 

In numerical studies in Section \ref{sec:simulation}, the proposed method is compared with three existing methods.  It should be mentioned that two of them, the dp-lasso and $t$lasso,  do not have the redescending property, and one of them, the nonparanormal, is not an M-estimator.  The redescending property for the dp-lasso and $t$lasso is discussed in Sections \ref{sec:dpd} and \ref{sec:tdistribution}.

\section{Algorithm}\label{sec:algorithm}

\citet{fujisawa2008robust} proposed an iterative minimization algorithm that monotonically decreases the negative $\gamma$-likelihood function $\ell_{\gamma}(\bm{\theta})$ given by (\ref{ガンマダイバージェンス}) at each step.  Their algorithm is based on the Pythagorean relation for the $\gamma$-divergence.  Unfortunately, their idea cannot be directly applied to our minimization problem of the penalized negative $\gamma$-likelihood $\ell_{\gamma,\lambda}(\bm{\theta})$ given by (\ref{罰則付きガンマダイバージェンス}), due to the $L_1$ penalization.  In this section, we propose an efficient minimization algorithm using the Majorize-Minimization algorithm (MM algorithm; \citealp{hunter2004tutorial,lange2010numerical}).

\subsection{Construction of the majorization function}
Let $\bm{\theta}^{(t)}$ be the estimate at the $t$th step.  We construct a majorization function of $\ell_1(\bm{\theta})$, say $\tilde{\ell}_1(\bm{\theta}|\bm{\theta}^{(t)})$.  The majorization function must satisfy the following properties:
\begin{eqnarray}
\tilde{\ell}_1(\bm{\theta}|\bm{\theta}^{(t)}) &\ge& \ell_1(\bm{\theta}),\label{MM_p1}\\
\tilde{\ell}_1(\bm{\theta}^{(t)}|\bm{\theta}^{(t)}) &=& \ell_1(\bm{\theta}^{(t)}).\label{MM_p2}
\end{eqnarray}
To construct a majorization function, first, we apply the Jensen's inequality to the convex function $y = -\log x$ and we have
\begin{eqnarray}
-\log\left(\sum_{i=1}^nw^{(t)}_i r_i^{(t)}\right) \le -\sum_{i=1}^n w^{(t)}_i\log r_i^{(t)}, \label{jensen}
\end{eqnarray}
where 
\begin{eqnarray}
w_i^{(t)} &=& \frac{f(\bm{x}_i;\bm{\theta}^{(t)})^{\gamma} }{\sum_{j=1}^nf(\bm{x}_j;\bm{\theta}^{(t)})^{\gamma}} 
= \dfrac{\exp \left\{-\frac{\gamma}{2} (\bm{x}_i - \bm{\mu}^{(t)})^T \bm{\Omega}^{(t)}(\bm{x}_i - \bm{\mu}^{(t)}) \right\}}{\sum_{j=1}^n \exp \left\{-\frac{\gamma}{2} (\bm{x}_j - \bm{\mu}^{(t)})^T \bm{\Omega}^{(t)}(\bm{x}_j - \bm{\mu}^{(t)}) \right\}}, \label{weight_gamma}\\
r_i^{(t)} &=& \sum_{j=1}^n f(\bm{x}_j;\bm{\theta}^{(t)})^{\gamma} \dfrac{ f(\bm{x}_i;\bm{\theta})^{\gamma}}{ f(\bm{x}_i;\bm{\theta}^{(t)})^{\gamma}}.\label{ri}
\end{eqnarray}
Here $\bm{\mu}^{(t)}$ and $\bm{\Omega}^{(t)}$ are the estimates of $\bm{\mu}$ and $\bm{\Omega}$ at the $t$th step, respectively.  Note that $\sum_{i=1}^n w_i^{(t)} = 1$ and $w_i^{(t)} r_i^{(t)} =  f(\bm{x}_i;\bm{\theta})^{\gamma}$.  Then, substituting (\ref{weight_gamma}) and (\ref{ri}) into (\ref{jensen}) gives us
\begin{eqnarray}
\ell_1(\bm{\theta}) 
\le - \sum_{i=1}^n  w_i^{(t)} \log f(\bm{x}_i;\bm{\theta}) + C, 
\label{MM1}
\end{eqnarray}
where $C = \frac{1}{\gamma} \sum_i w_i^{(t)}\log w_i^{(t)} + \frac{1}{\gamma}\log n$.  
Finally, we define  $\tilde{\ell}_1(\bm{\theta}|\bm{\theta}^{(t)})$ as the right side of (\ref{MM1}), i.e., 
\begin{equation}
\tilde{\ell}_1(\bm{\theta}|\bm{\theta}^{(t)}) = -  \sum_{i=1}^n  w_i^{(t)} \log f(\bm{x}_i;\bm{\theta}) + C. \label{l1tilde}
\end{equation}
It is shown that the function (\ref{l1tilde}) satisfies the properties of (\ref{MM_p1}) and (\ref{MM_p2}).

The majorization function (\ref{l1tilde}) is viewed as a weighted negative log-likelihood function with weights $w_i^{(t)}$ ($i=1,\dots,n$).  When $\bm{x}_i$ is an outlier, the corresponding likelihood $f(\bm{x}_i;\bm{\theta}^{(t)})$ is expected to be sufficiently small, so that the weight $w_i^{(t)}$, which is proportional to the $\gamma$th power of likelihood, is expected to be sufficiently small.

\subsection{Update algorithm}
We propose an update algorithm given by
\begin{eqnarray*}
\bm{\theta}^{(t+1)} = \arg\min_{\bm{\theta}} \tilde{\ell}_{\gamma,\lambda}(\bm{\theta}|\bm{\theta}^{(t)}),
\end{eqnarray*}
where
\begin{eqnarray*}
\tilde{\ell}_{\gamma,\lambda}(\bm{\theta}|\bm{\theta}^{(t)}) =  \tilde{\ell}_1(\bm{\theta}|\bm{\theta}^{(t)}) + \ell_2(\bm{\theta})  + \frac{\lambda}{2}  \| \bm{\Omega} - {\rm diag}(\bm{\Omega}) \|_1.
\end{eqnarray*}
From the properties of the majorization function given by (\ref{MM_p1}) and (\ref{MM_p2}), the target function $\ell_{\gamma,\lambda}(\bm{\theta})$ in (\ref{罰則付きガンマダイバージェンス}) monotonically decreases at each step: $\ell_{\gamma,\lambda}(\bm{\theta}^{(t)}) \ge \ell_{\gamma,\lambda}(\bm{\theta}^{(t+1)})$.

After a simple calculation of $\ell_2(\bm{\theta})$, which is detailed in Appendix A, we have 
\begin{eqnarray*}
\tilde{\ell}_1(\bm{\theta}|\bm{\theta}^{(t)}) + \ell_2(\bm{\theta}) = -\frac{1}{2(1+\gamma)}  \log|\bm{\Omega}| + \frac{1}{2}  {\rm tr} \left( \bm{\Omega} \bm{S}_{w^{(t)}} (\bm{\mu}) \right) +C',
\end{eqnarray*}
where
\begin{eqnarray*}
\bm{S}_{w^{(t)}}(\bm{\mu}) = \sum_{i=1}^n w_i^{(t)}(\bm{x}_i - \bm{\mu})(\bm{x}_i - \bm{\mu})^T,
\end{eqnarray*}
and $C'$ is a constant.  
For any $\bm{\Omega}$, the majorization function $\tilde{\ell}_{\gamma,\lambda}(\bm{\theta}|\bm{\theta}^{(t)})$ is minimized at
\begin{eqnarray*}
\bm{\mu}^{(t+1)}  = \sum_{i=1}^n w_i^{(t)}\bm{x}_i.
\end{eqnarray*}  For given  $\bm{\mu}^{(t+1)}$, the inverse covariance matrix $\bm{\Omega}^{(t+1)}$ is obtained by minimizing the following function with respect to  $\bm{\Omega}$:
\begin{eqnarray}
-\frac{1}{2(1+\gamma)}  \log|\bm{\Omega}| + \frac{1}{2}  {\rm tr} \left( \bm{\Omega} \bm{S}_{w^{(t)}}(\bm{\mu}^{(t+1)})   \right)   + \frac{\lambda}{2} \| \bm{\Omega} - {\rm diag}\bm{\Omega} \|_1.\label{dgammatilde}
\end{eqnarray}
The above minimization problem corresponds to the graphical lasso in (\ref{解きたい問題_adj}).  We can use a standard algorithm of the graphical lasso, such as the blockwise coordinate descent algorithm \citep{friedman2008sparse}, to obtain $\bm{\Omega}^{(t+1)}$.  

\begin{Remark}
	When $\gamma=0$,  the MM algorithm corresponds to the standard graphical lasso.  
\end{Remark}
\begin{Remark}
When $\lambda=0$, our update algorithm is identical to that of Example 4.1 in \citet{fujisawa2008robust}, in which the estimation algorithm is constructed by using the Pythagorean relation for the $\gamma$-divergence.
\end{Remark}

\subsection{Computation of entire path of solutions}
In practice, we set a sequence of decreasing regularization parameters $\lambda_1,\dots,\lambda_K$ on the log scale, and an entire path of solutions is made by sequences of $\lambda$.  The determination of the value of $\lambda_1$, which is a minimum value of $\lambda$ so that all of the non-diagonal elements of inverse covariance matrix are zeros, is provided in Appendix B. $\lambda_K$ is determined by $\lambda_K = \delta \lambda_1$, where $\delta$ is a positive value smaller than 1.    In our {\tt R} package {\tt rsggm}, the default is $\delta=0.2$ and $K=10$.

\section{Comparison with existing methods}\label{sec:comparison}
\subsection{dp-lasso}\label{sec:dpd}
\citet{sun2012robust} considered the problem of minimizing the following objective function based on the density power divergence \citep{basu1998robust}: 
\begin{eqnarray}
\ell_{\beta,\lambda}(\bm{\theta}) = \ell_{\beta}(\bm{\theta})  +  \frac{\lambda}{2}  \| \bm{\Omega} - {\rm diag}(\bm{\Omega}) \|_1,  \label{βダイバージェンス}
\end{eqnarray}
where $\ell_{\beta}(\bm{\theta})  $ is the modified likelihood function based on the density power divergence given by
\begin{eqnarray}
\ell_{\beta}(\bm{\theta}) = -\frac{1}{n\beta}   \sum_{i=1}^n f(\bm{x}_i;\bm{\theta})^{\beta}  + \frac{1}{1+\beta}  \int f(\bm{x};\bm{\theta})^{1+\beta}d\bm{x}.  \label{βダイバージェンス_罰則なし}
\end{eqnarray}
Here $\beta \ge 0$ controls the balance between efficiency and robustness.  \citet{miyamura2006robust} and \citet{sun2012robust} used $\ell_{\beta}(\bm{\theta})$ for robust estimation of the Gaussian graphical models.  We call the minimization problem of (\ref{βダイバージェンス}) the {\it dp-lasso}.

The density power divergence is quite similar to the $\gamma$-divergence.  The difference between $\ell_{\gamma}(\bm{\theta})$ and $\ell_{\beta}(\bm{\theta})$ is just the existence of the logarithm on $\ell_{\gamma}(\bm{\theta})$.  However, the estimators based on these two likelihood functions have much different robustness properties: the $\gamma$-lasso has the redescending property but the dp-lasso does not.  In fact, the kernel function of estimating equation of the dp-lasso can be expressed as
\begin{eqnarray}
\vpsi(\vx;\vtheta) = - f(\vx;\vtheta)^\gamma \vs(\vx;\vtheta) + \frac{\partial}{\partial\vtheta} b_{\beta}(\bm{\theta}) + \frac{\lambda}{2} \bm{u},\label{eq:estimating dp-lasso}
\end{eqnarray}
where 
$$
 b_{\beta}(\bm{\theta}) = \frac{|\bm{\Omega}|^{\beta/2}}{(1+\beta)^{1+p/2} (2\pi)^{p\beta/2}}.
$$
The first term of right side of (\ref{eq:estimating dp-lasso}) becomes zero as $\|\vx\| \rightarrow \infty$.  However, we have 
$\frac{\partial}{\partial\omega_{jk}}  b_{\beta}(\bm{\theta})$ does not always converges to 0 as $\|\vx\| \rightarrow \infty$, and therefore $\vpsi(\vx;\vtheta)$ does not go to zero as $\|\vx\| \rightarrow \infty$.

For the $\gamma$-divergence, we proposed the MM algorithm by using the concavity of the function $f(x) = \log x$, which  guarantees that the objective function monotonically decreases at each step.  On the other hand, it is difficult to derive the MM algorithm for the density power divergence.  \citet{sun2012robust} applied the coordinate descent algorithm using a quadratic approximation \citep{tseng2009coordinate}, but it does not guarantee that the objective function monotonically decreases at each step.  Furthermore, their algorithm has additional difficulties to be applied as follows:
\begin{itemize}
\item Their algorithm does not guarantee to have a positive definite inverse covariance matrix. 
\item Their algorithm based on \citet{tseng2009coordinate} depends on four tuning parameters to be determined.  
\end{itemize}
In the $\gamma$-lasso, the estimated inverse covariance matrix is positive definite, and the update algorithm does not have any tuning parameter to be determined in advance.

\subsection{\tlasso}\label{sec:tdistribution}
\citet{finegold2011robust} proposed a penalized maximum likelihood approach using a multivariate $t$-distribution instead of a Gaussian distribution, which was referred to as the {\it \tlasso}.  The density function of the multivariate $t$-distribution with mean vector $\bm{\mu}$, shape matrix $\bm{\Sigma}$, and the degrees of freedom $\nu$ is 
\begin{eqnarray*}
f(\bm{x};\bm{\mu},\bm{\Sigma},\nu) = \frac{\Gamma(\frac{\nu +p}{2})|\bm{\Sigma}|^{-1/2}} {(\pi\nu)^{1/2}\Gamma(\frac{\nu}{2}) \{ 1+ (\bm{x} - \bm{\mu})^T\bm{\Sigma}^{-1}(\bm{x} - \bm{\mu})/\nu \}^{(p+\nu)/2}}.
\end{eqnarray*}
With a simple calculation of the EM algorithm \citep{finegold2011robust}, we can obtain the following iterative algorithm: 
\begin{eqnarray*}
\bm{\mu}^{(t+1)} &=&  \sum_{i=1}^n {u}_i^{(t)}\bm{x}_i \\
\bm{\Omega}^{(t+1)} &=& {\rm arg} \min_{\bm{\Omega}} \left\{-  \log|\bm{\Omega}| +  {\rm tr} \left( \bm{\Omega} \bm{S}_{u_i^{(t)}} (\bm{\mu}^{(t+1)}) \right)  + \lambda\| \bm{\Omega} - {\rm diag}\bm{\Omega} \|_1 \right\},
\end{eqnarray*}
where
\begin{eqnarray*}
\bm{S}_{u^{(t)}}(\bm{\mu}) = \sum_{i=1}^n u_i^{(t)} (\bm{x}_i - \bm{\mu})(\bm{x}_i - \bm{\mu})^T,
\end{eqnarray*}
and
\begin{eqnarray*}
u_i^{(t)} = \frac{\tilde{u}_i^{(t)}}{\sum_{j=1}^n \tilde{u}_j^{(t)}}, \quad \tilde{u}_i^{(t)} = \frac{\nu+p}{\nu + (\bm{x}_i - \bm{\mu}^{(t)})^T \bm{\Omega}^{(t)}(\bm{x}_i - \bm{\mu}^{(t)})}.
\end{eqnarray*}
The EM algorithm in the multivariate $t$-distribution turns out to be the problem of the weighted graphical lasso.  

We recall that the $\gamma$-lasso is also based on the weighted graphical lasso.  However, there are two significant differences between the EM algorithm and our algorithm as follows:
\begin{itemize}
	\item In the EM algorithm, the first term of the complete-data log-likelihood function is $\frac{1}{2} \log \bm{\Omega}$, whereas in the \gmethod, the corresponding term is $\frac{1}{2(1+\gamma)} \log \bm{\Omega}$. 
	\item The formula of the weight implies that the $\gamma$-divergence is more robust than the multivariate $t$-distribution.  For example, suppose that the $i$th observation is an outlier. As $\|\bm{x}_i\| \rightarrow \infty$, we expect that the estimator is not affected by $\bm{x}_i$.  In the EM algorithm, although $u_i^{(t)} \rightarrow 0$ as $\|\bm{x}_i\| \rightarrow \infty$, ${u}_i^{(t)}(\bm{x}_i-\hat{\bm{\mu}}^{(t)})(\bm{x}_i-\hat{\bm{\mu}}^{(t)})^T$ does not become zero as $\|\bm{x}_i\| \rightarrow \infty$.  This means the $\bm{S}_{u^{(t)}}(\hat{\bm{\mu}}^{(t)})$ is sensitive to the outlier when $\|\bm{x}_i\|$ is large.  On the other hand, for the $\gamma$-divergence, ${w}_i^{(t)}(\bm{x}_i-\hat{\bm{\mu}}^{(t)})(\bm{x}_i-\hat{\bm{\mu}}^{(t)})^T$ becomes zero as $\|\bm{x}_i\| \rightarrow \infty$.     
\end{itemize}

It is shown the $t$lasso does not have the redescending property. The kernel function of the estimating equation for $t$lasso is given by
\begin{eqnarray*}
\vpsi(\vx;\vtheta) = - \frac{\partial }{\partial\vtheta} \log f(\bm{x};\bm{\mu},\bm{\Sigma},\nu) +  \frac{\lambda}{2}\bm{u}.
\end{eqnarray*}
In general, $\frac{\partial}{\partial\omega_{jk}}  \log f(\bm{x};\bm{\mu},\bm{\Sigma},\nu) \neq 0$ as $\|\bm{x}\| \rightarrow \infty$, so that $\vpsi(\vx;\vtheta)$ does not go to zero as $\|\bm{x}\| \rightarrow \infty$.

\subsection{Nonparanormal}
\citet{liu2009nonparanormal} proposed the {\it nonparanormal}, which uses the semiparametric Gaussian copula for estimating the graph.  The nonparanormal allows the transformation of non-normal data to normal data, which enables us to weaken the assumption of normality.  Let $h$ be a monotone and differentiable function such that $h(\bm{X}) = (h(X_1),\dots,h(X_p))^T$ is multivariate-normally distributed with mean vector $\bm{\mu}$ and the covariance matrix $\bm{\Omega}^{-1}$.   \citet{liu2009nonparanormal} showed that when $h_j(x) = \mu_j + \sqrt{\sigma_{jj}} \Phi^{-1}(F_j(x))$, $X_i$ and $X_j$ is conditionally independent if and only if $\omega_{ij}=0$, where $F_j(x)$ is the cumulative distribution function (CDF) of the marginal distribution of $h(\bm{X})$, $\Phi$ is the CDF of the standard normal distribution, and $\sigma_{jj}$ is the $(j,j)$th element of $\bm{\Omega}^{-1}$.  \citet{liu2009nonparanormal} estimated $h_j(x)$ by $\hat{h}_j(x) = \hat{\mu}_j + \sqrt{\hat{\sigma}
 _{jj}} \Phi^{-1}(\tilde{F}_j(x))$, where $\hat{\mu}_j$ is the sample mean, $\sqrt{\hat{\sigma}_{jj}}$ is the sample standard deviation, and $\tilde{F}_j$ is the truncated empirical CDF defined as  
\begin{eqnarray*}
	\tilde{F}_j(x) = 
\left\{
\begin{array}{ll}
\delta_n & \mbox{if } \hat{F}_j(x) < \delta_n\\
\hat{F}_j(x) & \mbox{if } \delta_n \le \hat{F}_j(x) \le 1-\delta_n\\
1-\delta_n & \mbox{if } \hat{F}_j(x) > 1-\delta_n\\
\end{array} 
\right..
\end{eqnarray*}
Here $ \hat{F}_j(x)$ is the empirical CDF and $\delta_n$ is a truncation parameter. \citet{liu2009nonparanormal} selected $\delta_n = \frac{1}{4n^{1/4}\sqrt{\pi\log n}}$ to achieve the desired rate of convergence in high-dimensional settings.

However, the truncation parameter selected by \citet{liu2009nonparanormal} may not appropriately treat the outliers:

\begin{itemize}
	\item We have $ \lim_{n \rightarrow \infty} \delta_n = 0$, which implies $\delta_n$ is too small to detect the outliers for large samples when the contamination ratio is large. 
	\item The nonparanormal cannot appropriately detect the outliers when the outliers are present only on one side, because the truncation is symmetric.  In fact, our simulation study presented in Section \ref{sec:simulation} also showed that the nonparanormal did not perform well when the outliers were present on one side.
\end{itemize}

Our proposed procedure, \gmethod, does not have any truncation parameter, which implies the \gmethod\ does not have an issue as above.  Furthermore, the \gmethod\ can appropriately treat the outliers even if the contamination ratio is large and/or the outliers are present only on one side, because the weight (\ref{weight_gamma}) of the outlier is expected to be sufficiently small.

\section{Simulation study}\label{sec:simulation}
\subsection{Simulation model}
In this simulation study, we used the following three simulation models:
\begin{eqnarray*}
{\rm (i)} & & (1-\ee) N_p(\bm{0},\bm{\Omega}^{-1})+\ee N_p(\bm{0},30 \bm{I}), \\
{\rm (ii)} & & (1-\ee) N_p(\bm{0},\bm{\Omega}^{-1})+\ee N_p(\eta \bm{1}, \bm{I}), \\
{\rm (iii)} & & (1-\ee) N_p(\bm{0},\bm{\Omega}^{-1})+\ee N_p(\eta \bm{1}^{(20)}, \bm{I}),
\end{eqnarray*}
where $\bm{I}$ is the identity matrix, $\bm{1}$ is the $p$-dimensional vector whose elements are one, $\bm{1}^{(20)}$ is the $p$-dimensional vector whose first 20 elements are one and latter $p-20$ elements are zeros, and $\varepsilon$ ($0\le\varepsilon<1$) is the contamination ratio.  The number of variables was set to be $p=100$.  For the model (i), the distribution of outliers are symmetric but away from the central tendency.  For the models (ii) and (iii), the outliers are present on one side of the mean direction of $\eta \bm{1}$ or $\eta \bm{1}^{(20)}$.

We generated the inverse covariance matrix $\bm{\Omega}$ in a manner similar to \citet{JMLR:v15:tan14b}. First, we generated an adjacency matrix $\bm{A} = (A_{ij})$ by the Barab\'asi-Albert model \citep{barabasi1999emergence}. Note that the degree distribution of the network generated by the Barab\'asi-Albert model follows power-law.  Many real-world networks are often considered as the scale-free networks, in which the degree distribution follows power-law \citep{barabasi1999emergence}.  Next, we created a matrix $\bm{E}=(E_{ij})$ given by  
\begin{equation*}
E_{ij} = \left\{ 
\begin{array}{ll}
U[{\mathcal D}] & \mbox{ if } A_{ij} = 1\\
0 & \mbox{ otherwise}\\
\end{array}
\right.
,
\end{equation*}
where $U[{\mathcal D}]$ is a random sample from a uniform distribution with ${\mathcal D} = [-0.75,-0.25] \cup [0.25,0.75]$.  We calculated $\tilde{\bm{E}} := (\bm{E} + \bm{E}^T) / 2 $ and set $\tilde{\bm{\Omega}} = \tilde{\bm{E}} + (0.1 - \lambda_{\min})\bm{I}$, where $\lambda_{\min}$ is the smallest eigenvalue of $\tilde{\bm{E}}$.  Finally, we set $\bm{\Omega} = \bm{L}^{1/2}\tilde{\bm{\Omega}}\bm{L}^{1/2}$, where $\bm{L} = {\rm diag}(\bm{\Omega}^{-1})$.  This procedure guarantees the positive definiteness of the inverse covariance matrix $\bm{\Omega}$.

\subsection{Monte Carlo simulations}
We conducted Monte Carlo simulations to investigate the performance of the proposed procedure. The values of $\varepsilon$ and $\eta$ were set to be $\varepsilon =0$, $0.05,$ $0.1,$ $0.3$ and $\eta=5$, $10$, respectively.  We generated 100 datasets with $n=200,$ $2000,$ $20000$.  The tuning parameters were set to be $\gamma=0.05$, $\beta=0.05$, and $\nu = 1$.   In the dp-lasso procedure, we were not able to obtain the solutions for a few datasets because of the non-convergence of the mean vector.  We summarized the result without using such datasets for the dp-lasso.  We compared (i) ROC curves and (ii) mean squared errors (MSEs) based on the non-diagonal parameters of $\bm{\Omega}$.  

\subsubsection{ROC curve}
\begin{figure}[!t]
\begin{center}
    \includegraphics[width=140mm]{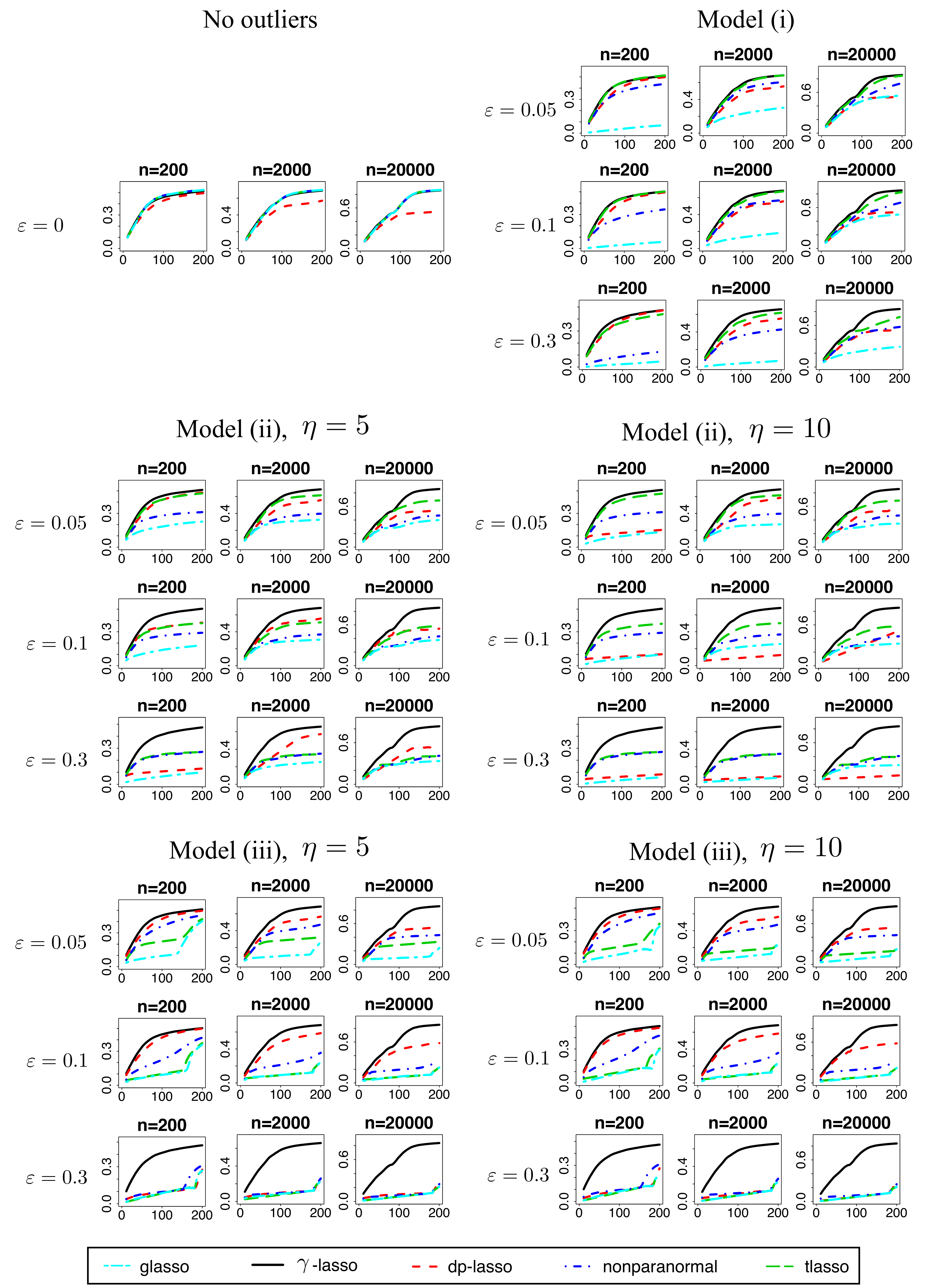}
\caption{ROC curves for simulated data. The $x$-axis indicates the number of estimated non-zero non-diagonal elements of $\bm{\Omega}$ and the $y$-axis is the mean of the true positive rate for the non-zero non-diagonal elements of $\bm{\Omega}$.}
\label{fig:ROC_simulation}
\end{center}
\end{figure}

The ROC curves were depicted in Figure~\ref{fig:ROC_simulation}. 
We obtain the following tendencies:
\begin{itemize}
	\item Clearly, the \gmethod\ significantly outperformed for most cases.  An important point is that the ROC curves for the \gmethod\ were essentially independent of contamination ratios, which implies that the performance of the \gmethod\ was stable.   On the other hand, the existing methods depended on the contamination ratios.  For example, in the model (iii), the existing methods performed much worse than the \gmethod\ for large contamination ratio.    
	\item The \tlasso\ showed a similar performance to the \gmethod\ in the model (i) in most cases.  However, the \tlasso\ was clearly worse than the \gmethod\ in the models (ii) as the contamination ratio became larger.  In the model (iii), the \tlasso\ showed a poor performance.  These poor performances may be caused by a symmetric multivariate $t$-distribution with a heavy tail. The symmetric distribution cannot appropriately treat the outliers when they are present on one side.
	\item The nonparanormal generally performed worse than the \gmethod\ in most cases, especially when the contamination ratio was large.   This will be because the truncation parameter of the truncated empirical distribution was selected by $\delta_n = \frac{1}{4n^{1/4}\sqrt{\pi\log n}}$; as $n \rightarrow \infty$, $\delta_n \rightarrow 0$, which suggests that the outliers may not be detected for large sample sizes (e.g. $\delta_n = 0.0038$ when $n = 20000$).   
	\item When $\varepsilon =0$, the dp-lasso performed worse than the other methods for large sample sizes.  When $n=200$, the dp-lasso sometimes showed a similar performance to the \gmethod. \end{itemize}

\subsubsection{MSE}\label{subsubsec:MSE}
\begin{figure}[!t]
\begin{center}
    \includegraphics[width=140mm]{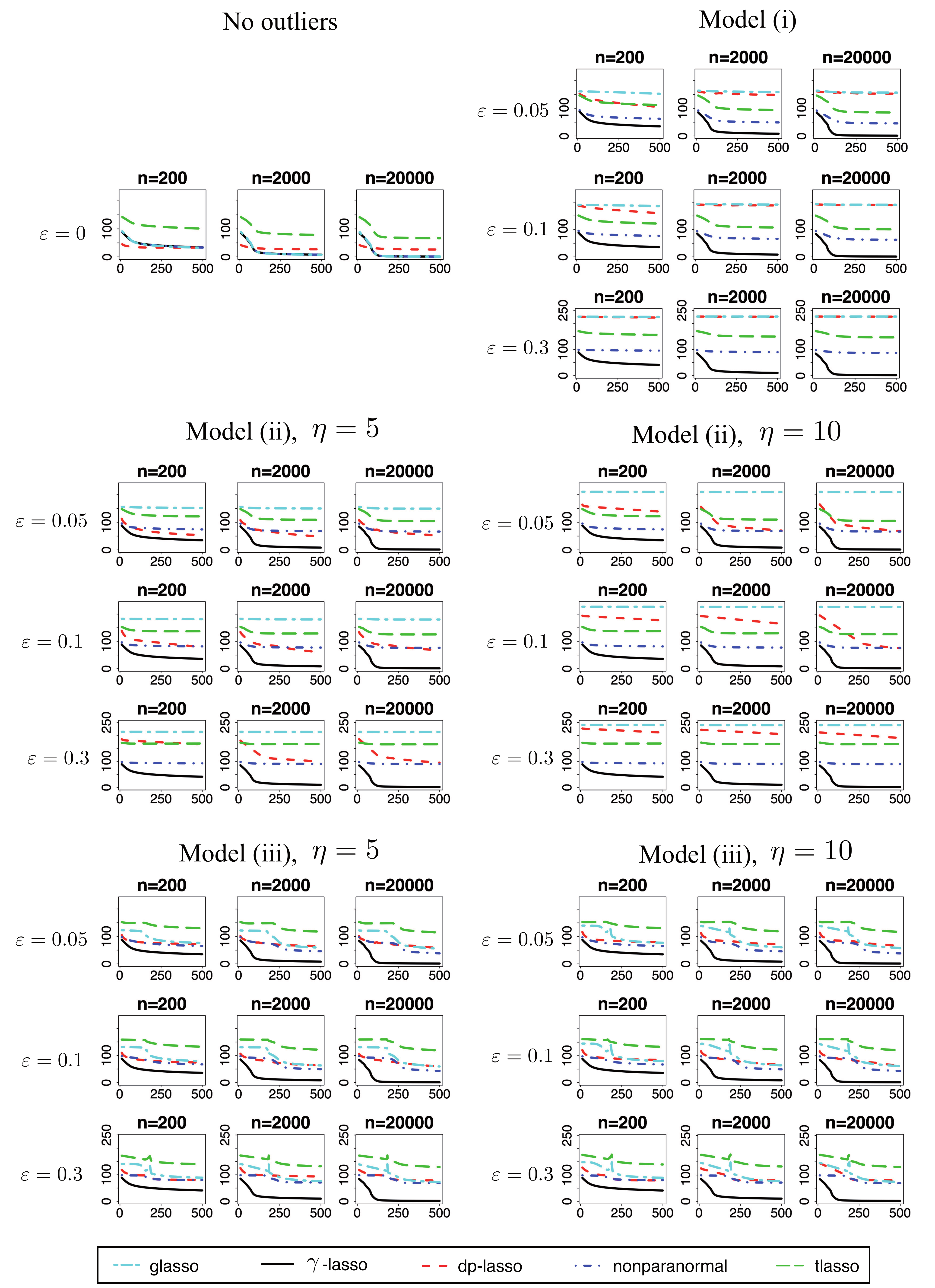}
\caption{MSEs for simulated data. The $x$-axis indicates the number of estimated non-zero non-diagonal elements of $\bm{\Omega}$ and the $y$-axis is the mean of MSE on the non-diagonal elements of $\bm{\Omega}$.}
\label{fig:MSE_simulation}
\end{center}
\end{figure}

The mean squared errors (MSEs) of the inverse covariance matrix $\bm{\Omega}$ are given in Figure~\ref{fig:MSE_simulation}. 
We obtain the following tendencies:
\begin{itemize}
\item The \gmethod\ clearly performed the best in most cases. In particular, when $n$ was large, the MSE approached zero as the estimated graph became dense (i.e., the value of $x$-axis in Figure \ref{fig:MSE_simulation} became large).  On the other hand, for the other methods, the MSE did not approach zero except for the case of  nonparanormal and the standard graphical lasso when $\varepsilon = 0$.  
\item The \tlasso\ performed well in terms of the ROC curve in the case (i), but did not perform well in terms of the MSE. This will be because the multivariate $t$-distribution with a heavy tail has a very large variance.  
\item The estimates of the dp-lasso and $t$lasso were biased even when $\varepsilon = 0$.  
\item The nonparanormal often showed the second smallest MSE. 
\end{itemize}

\section{Gene expression data analyses}
\subsection{Galactose utilization}\label{sec:real1}

The yeast gene expression data were provided by \citet{gasch2000genomic}.  We restrict our attention to 8 genes involved in galactose utilization \citep{ideker2001integrated}. \citet{finegold2011robust} reported that 11 out of the 136 experiments showed unusually large negative values for 4 out of these 8 genes: GAL1, GAL2, GAL7, GAL10. Using the dataset, \citet{vinciotti2013robust} compared several estimation procedures via $L_1$ penaization, including the standard graphical lasso,  the \tlasso, and the nonparanormal.  They also applied the graph estimation approach given by \citet{meinshausen2006high}, in which the lasso regression  \citep{Tibshirani:1996} is carried out to each variable.   \citet{vinciotti2013robust} applied not only the ordinary lasso regression, but also the adaptive lasso \citep{Zou:2006}, and two typical robust estimation procedures, Least Absolute Deviation and Huber function,
  with the weighted lasso. The data were normalized to have sample mean 0 and sample deviation 1 before applying the above methods.  The tuning parameter $\lambda$ was selected so that the number of edges was 9. 

\citet{vinciotti2013robust} estimated the edges using the original data and the uncontaminated data in which 11 outliers were removed from the original data. Let the original data and the uncontaminated data be denoted by $\bm{X}$ and $\bm{X}_{(-11)}$, respectively.  A set of edges estimated using $\bm{X}$ was not always the same as that estimated using $\bm{X}_{(-11)}$. The difference was examined by the {\it total agreement} defined as follows: Let $A$ and $B$ be the set of edges estimated using $\bm{X}$ and $\bm{X}_{(-11)}$, respectively. Let $A^c=C-A$ and $B^c=C-B$, where $C$ is a set of edges of the complete graph. Let $\# D$ be the number of elements in the set $D$. The total agreement is given by $\{\#(A \cap B) + \# (A^c \cap B^c) \}/\# C$.

\citet{vinciotti2013robust} reported that the total agreement of the estimated edges was at most 0.86 among various robust estimation procedures described as above.  We applied the \gmethod\ with $\gamma=0.05,$ $0.1,$ $0.5$ and dp-lasso with $\beta=0.05,$ $0.1,$ $0.5$, which had not been applied yet. The results are given in Table~\ref{table:real1}.    The \gmethod\ showed that the total agreements were 1 for any $\gamma$, which means that the graph estimated using $\bm{X}$ were completely same as that estimated using $\bm{X}_{(-11)}$. Therefore, the \gmethod\ was stable for any $\gamma$.  The total agreement of the dp-lasso was also 1 for $\beta = 0.5$, whereas less than 0.8 for $\beta = 0.1$ and $\beta = 0.05$.  Thus, the dp-lasso was sensitive to the tuning parameter $\beta$. 

We also examined another normalization of the data, because the sample mean and sample deviation are not robust to outliers. The data were normalized with the robust estimates of mean and scale (median and adjusted median absolute deviation (MAD)) before the robust analyses.  The results are also given in Table~\ref{table:real1}.   In most cases, the performances based on the MAD were better than those based on the SD.  Our procedure yielded the largest total agreement.   When the dp-lasso was applied with $\beta=0.5$, we were not able to find a tuning parameter $\lambda$ whose solution had 9 edges because of the instability of the solution path.  

\begin{figure}[!t]
\centering
\subfigure[graphical lasso, removing 11 observations]{\includegraphics[width=4.5cm]{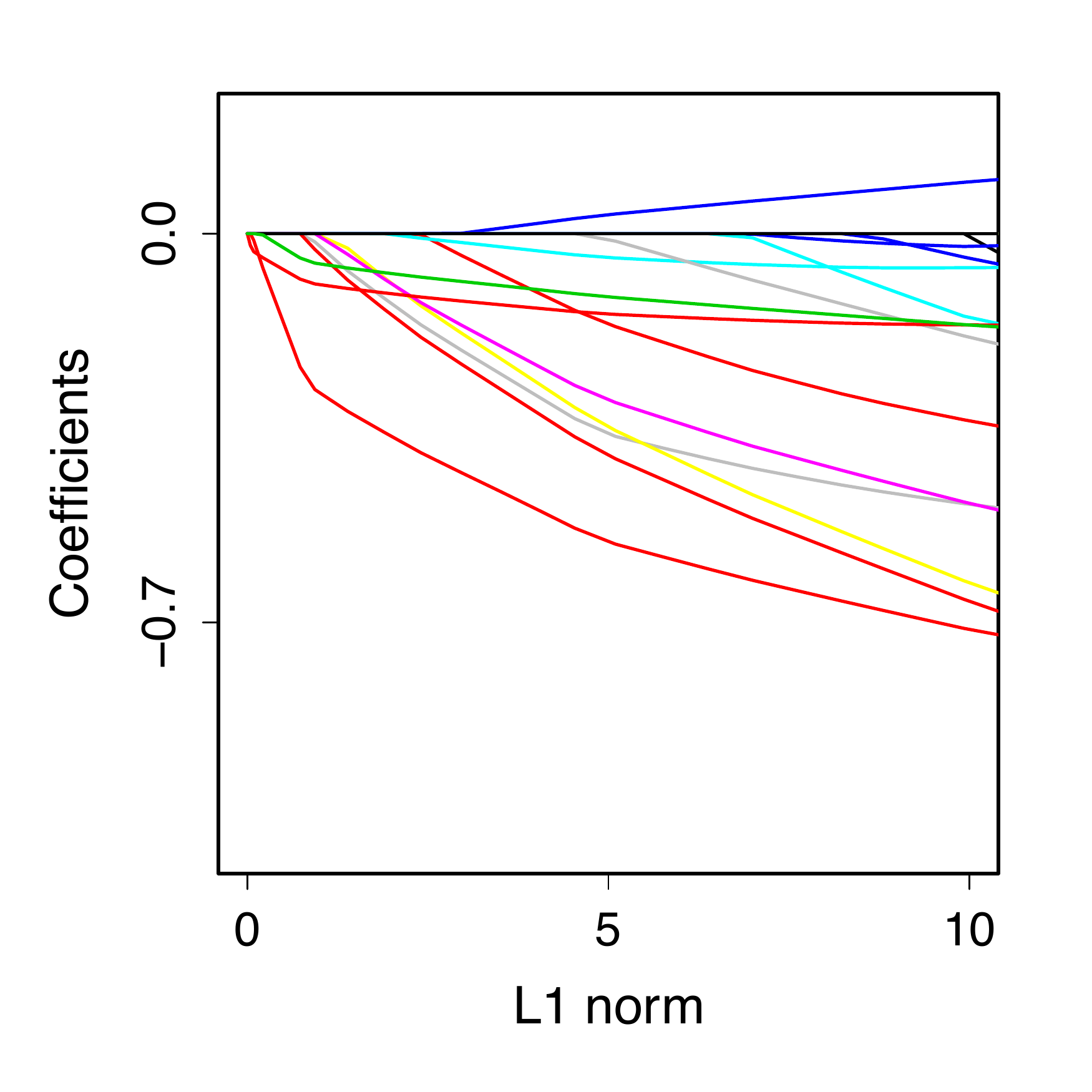}
 \label{fig:gasch_solpath_lasso_nooutlier}}
\subfigure[graphical lasso]{\includegraphics[width=4.5cm]{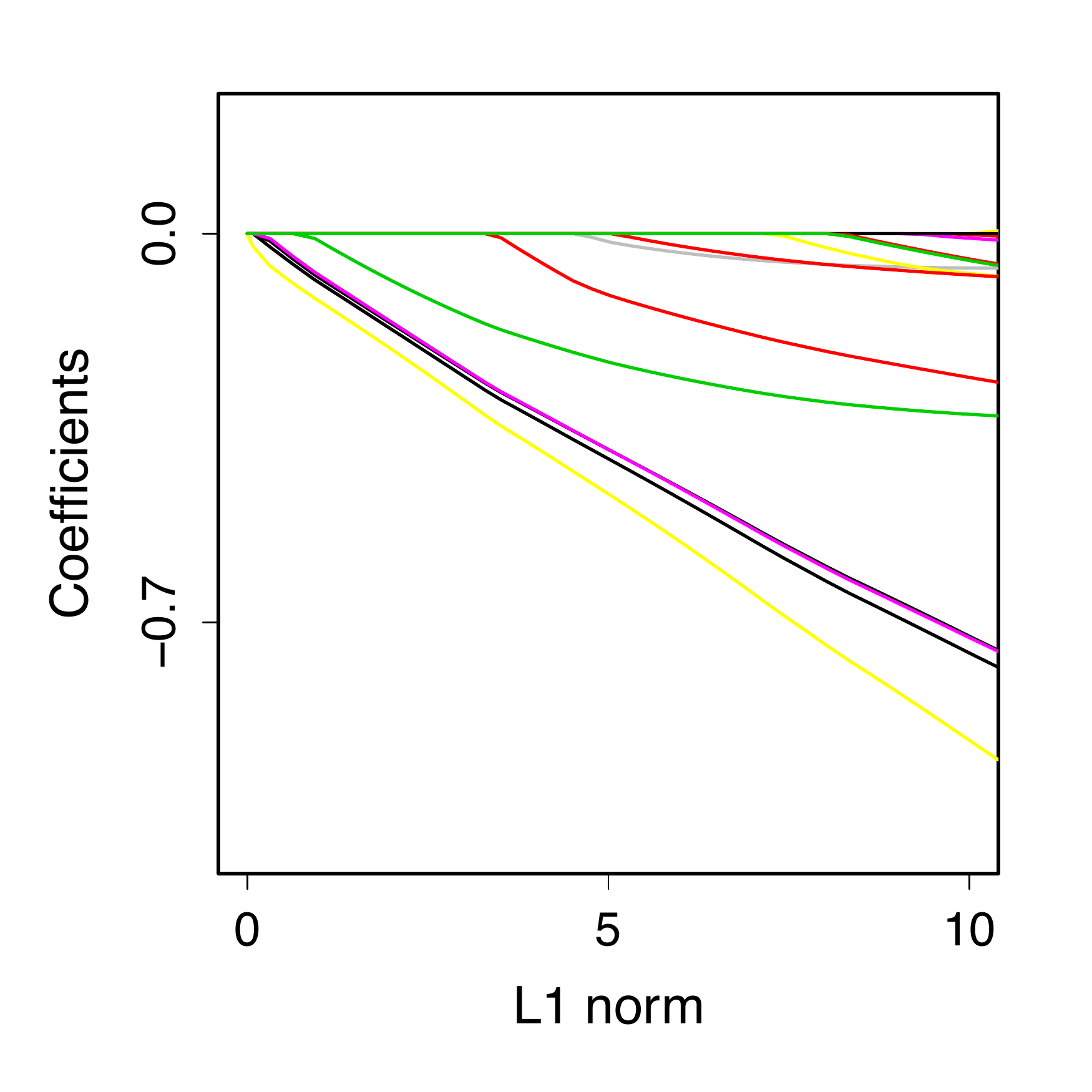}
 \label{fig:gasch_solpath_lasso}}
\subfigure[$\gamma$-lasso]{\includegraphics[width=4.5cm]{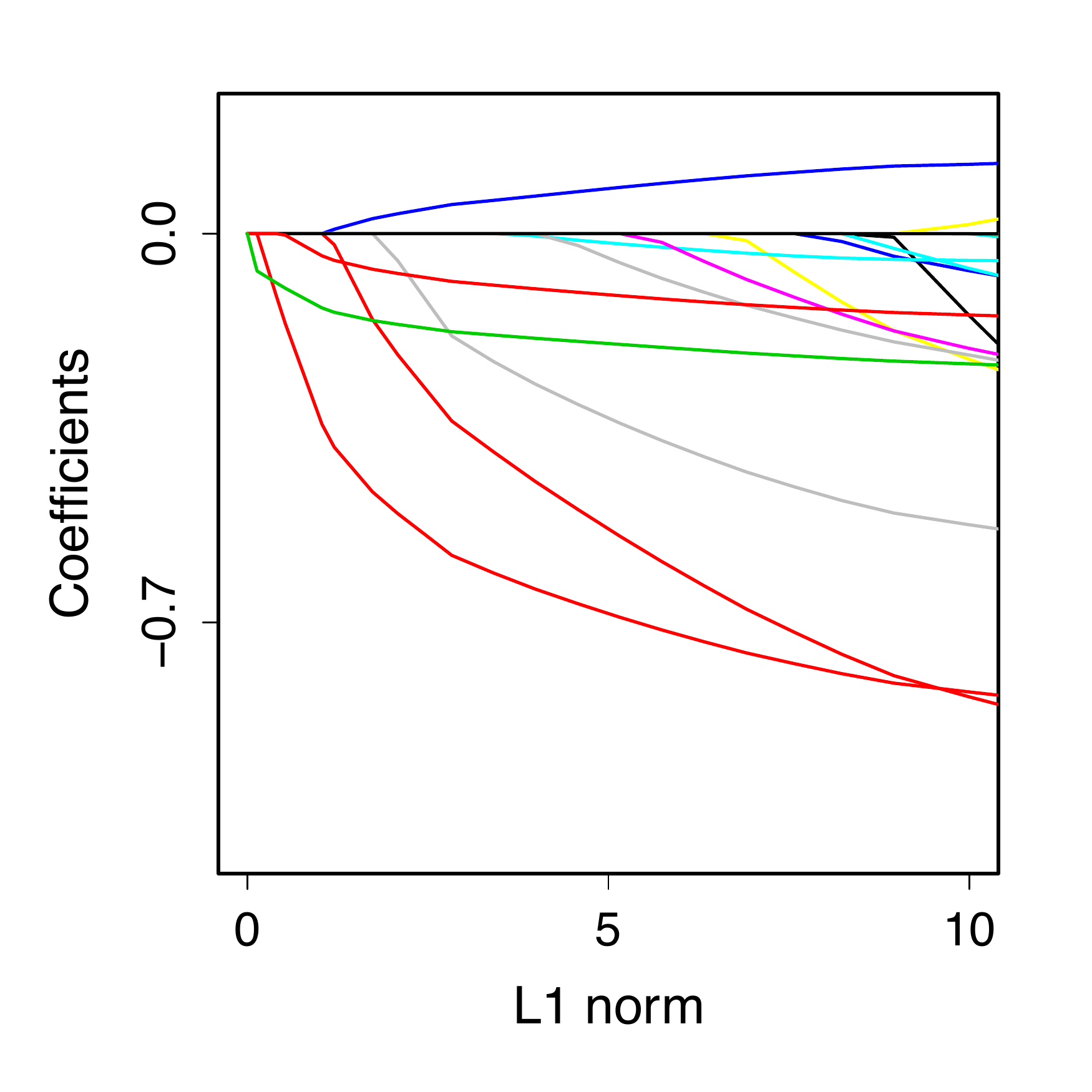}
 \label{fig:gasch_solpath_gamma-lasso}}
 \\
\subfigure[nonparanormal]{\includegraphics[width=4.5cm]{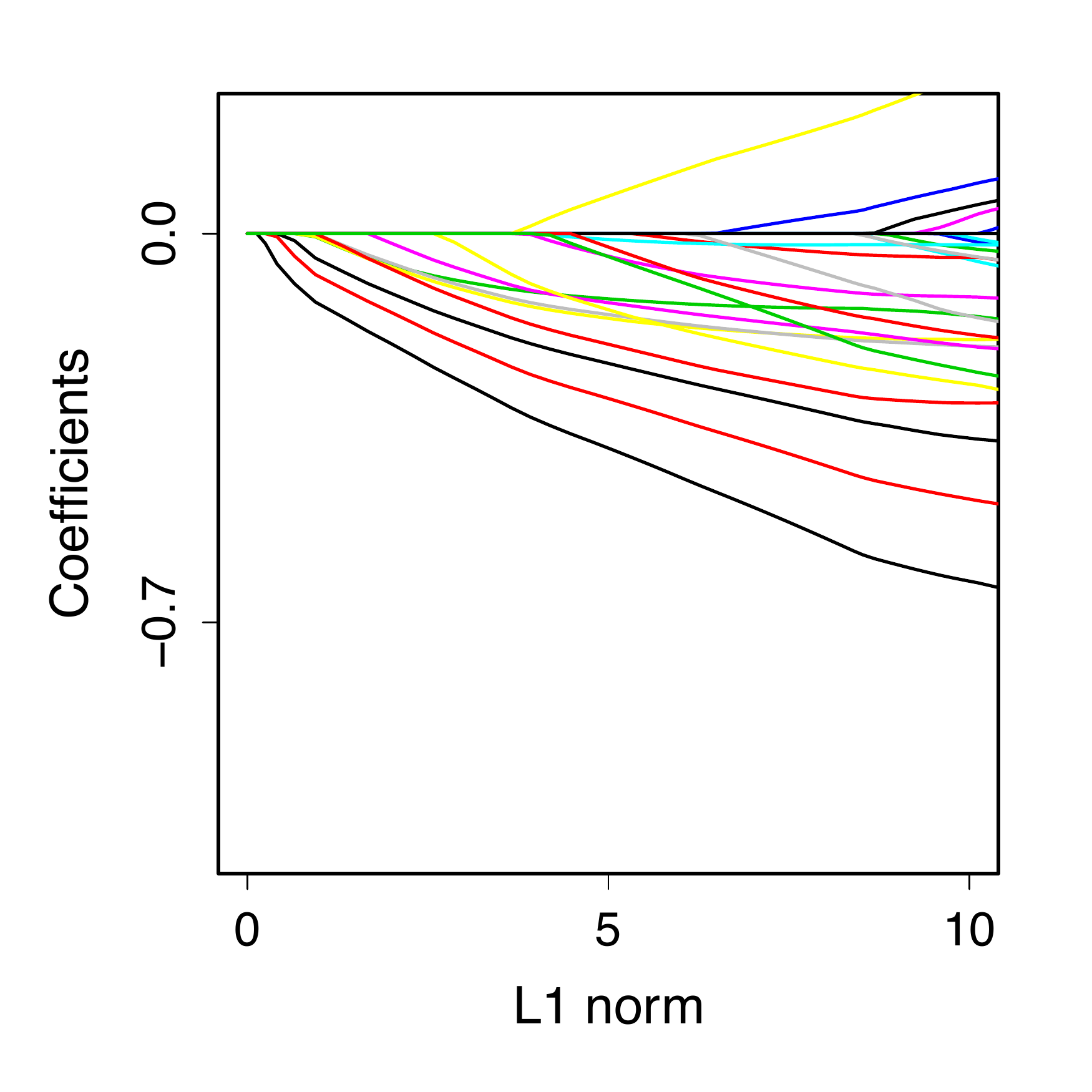}
 \label{fig:gasch_solpath_nonparanormal}}
\subfigure[tlasso]{\includegraphics[width=4.5cm]{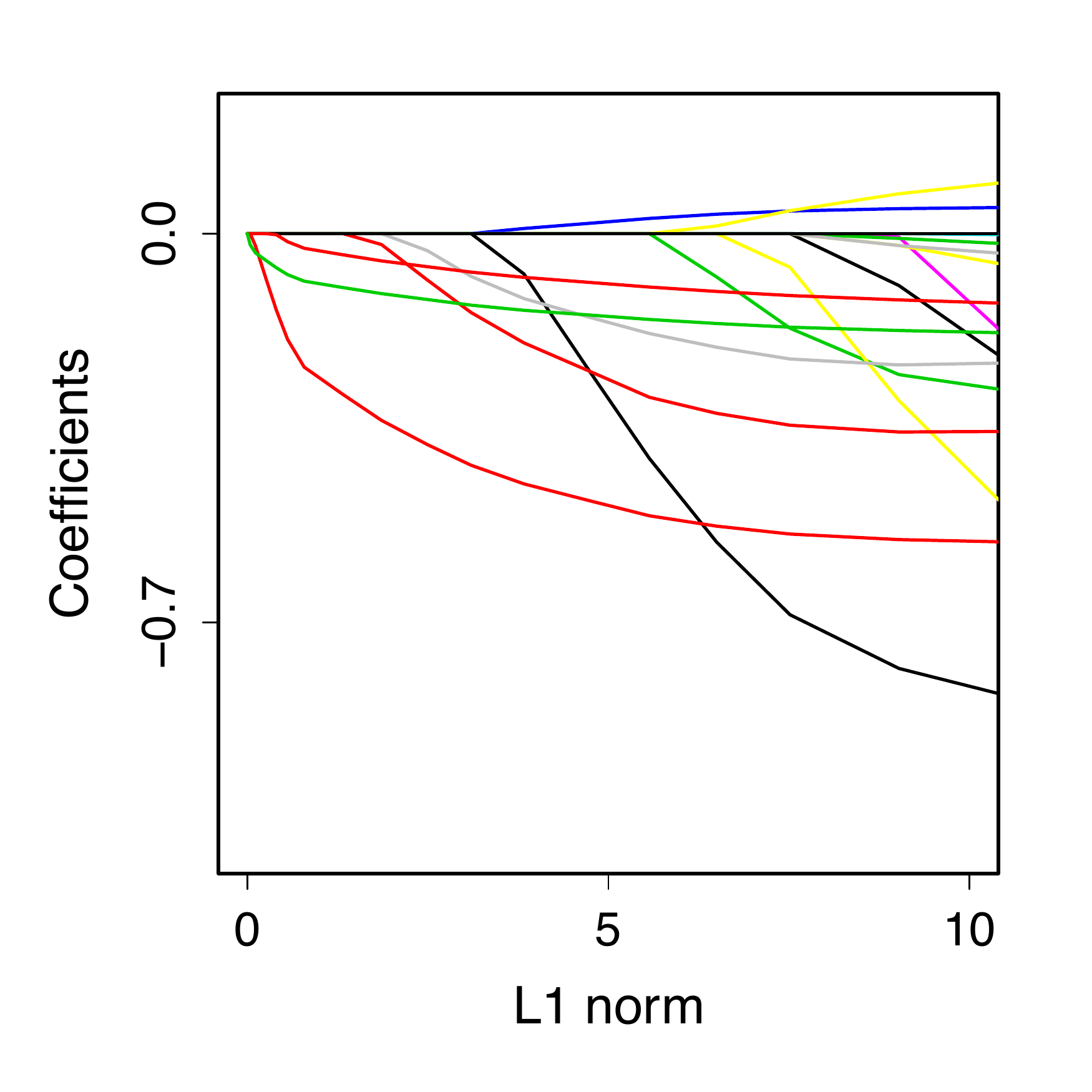}
 \label{fig:gasch_solpath_tlasso}}
\subfigure[dp-lasso]{\includegraphics[width=4.5cm]{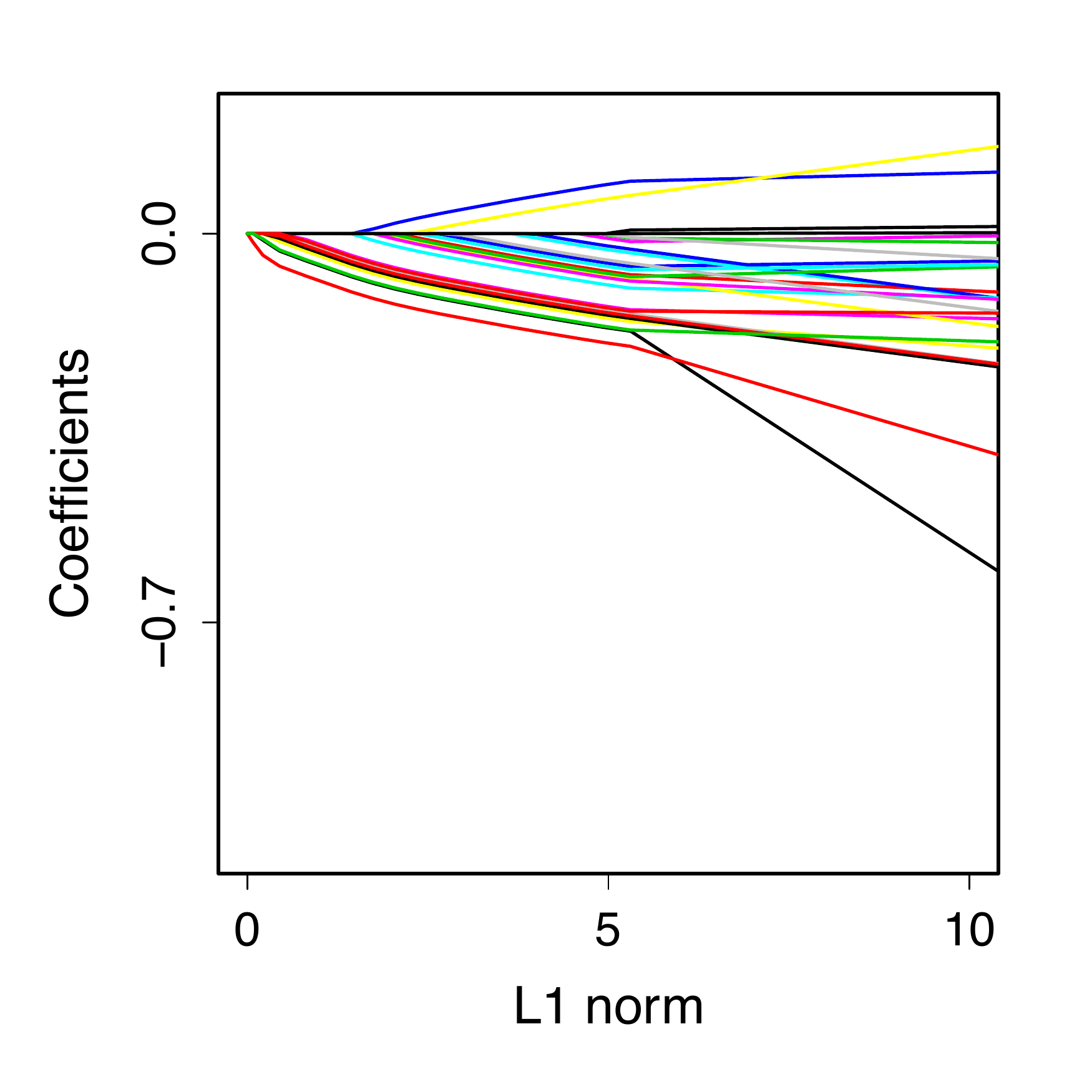}
 \label{fig:gasch_solpathdp-lasso}}
\subfigure[Weights obtained by the $\gamma$-lasso]{\includegraphics[width=9cm]{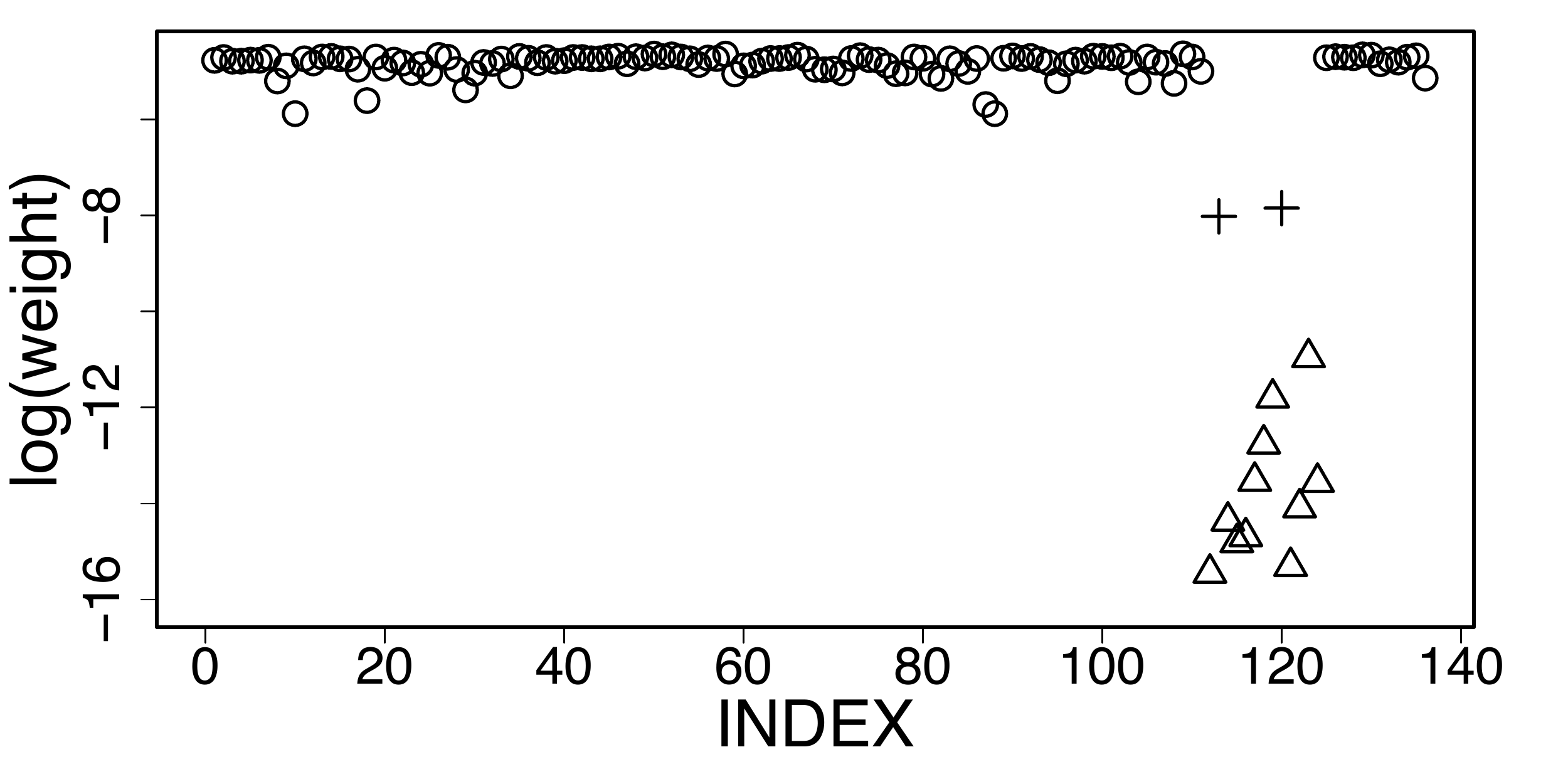}
 \label{fig:gasch_weight}}
\subfigure[graphical lasso, removing 13 observations]{\includegraphics[width=4.5cm]{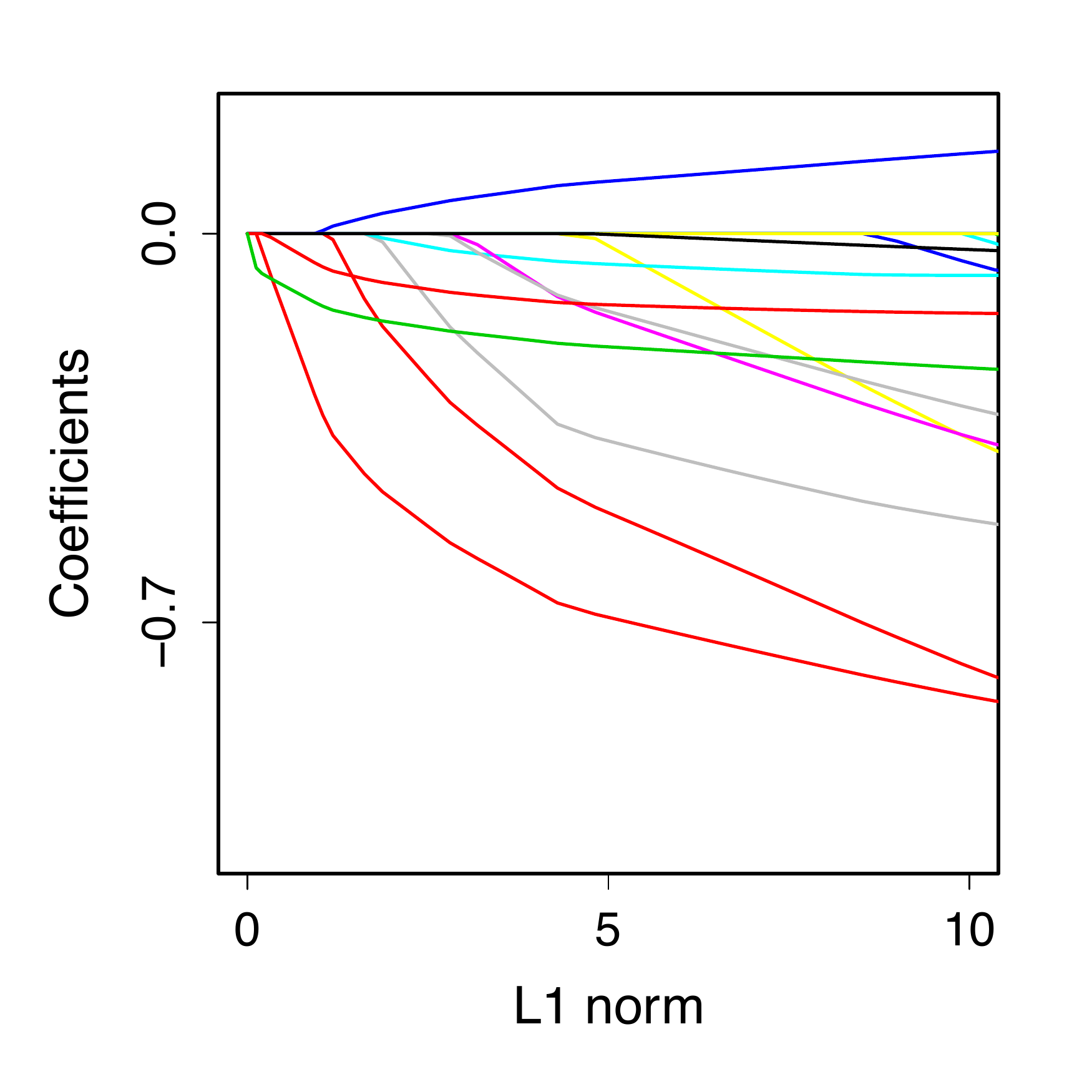}
 \label{fig:Gasch_solpathdp-lasso}}
\caption{The solution paths (estimates of $\omega_{ij}$ ($i,j=1,\dots, p$, $i<j$) as a function of $\sum_{i < j}|\hat{\omega}_{ij}|$) and the weights of the \gmethod\ in (\ref{weight_gamma}) for the yeast gene expression data. (a)  The solution path for the graphical lasso applied to $\bm{X}_{(-11)}$. (b)-(f) The solution paths made by the above methods applied to $\bm{X}$.  (g) The weight values of \gmethod\ given by (\ref{weight_gamma}).  We could find two additional data values whose weight values were enough small (denoted ``+") to be regarded as outliers in addition to the 11 outliers (denoted ``$\bigtriangleup$").  (h) The solution path of the graphical lasso applied to the data removing the 13 outliers (original 11 outliers and additional 2 outliers).  }
\label{fig:Gasch_solpath}
\end{figure}

We also depicted the solution paths of the non-diagonal elements of inverse covariance matrix, which are given in Figure~\ref{fig:Gasch_solpath}. Figure~\ref{fig:gasch_solpath_lasso_nooutlier} is the solution path for the graphical lasso applied to $\bm{X}_{(-11)}$, and the Figures \ref{fig:Gasch_solpath}(b)-(f) depict the solution paths of the various methods (standard graphical lasso, \gmethod, dp-lasso, $t$lasso, nonparanormal) applied to $\bm{X}$. Among Figures \ref{fig:Gasch_solpath}(b)-(f), only Figure \ref{fig:Gasch_solpath}(c) was similar to Figure \ref{fig:Gasch_solpath}(a).  This means the \gmethod\ was stable against the outliers, whereas the other methods were sensitive to the outliers. 

Figure~\ref{fig:gasch_weight} showed the weight values of \gmethod\ given by (\ref{weight_gamma}).  We could find two additional data values whose weight values were enough small to be regarded as outliers.  Figure~\ref{fig:Gasch_solpath}(h) depicts the solution path of the graphical lasso applied to the data removing the 13 outliers (original 11 outliers and additional 2 outliers).  Figure~\ref{fig:Gasch_solpath}(c) was more similar to Figure~\ref{fig:Gasch_solpath}(h) than Figure \ref{fig:Gasch_solpath}(a).  The \gmethod\ performed as if the 13 outliers were known in advance.

\begin{table}[!t]
\centering
 \caption{Comparison of Total Agreement and Common Edges.  The SD in left column indicates that the data were normalized to have sample mean 0 and sample deviation 1.  The MAD in the right column indicates that the data were normalized with the robust estimates of mean and scale (median and adjusted median absolute deviation (MAD)).} \label{table:real1}
\begin{tabular*}{16cm}{@{\extracolsep{\fill}}lrrrrr}
  \hline
 & \multicolumn{2}{c}{SD}  & \multicolumn{2}{c}{MAD}  \\ 
 & {\small Total Agreement} & {\small Common Edges} & {\small Total Agreement} & {\small Common Edges} \\ 
  \hline
glasso                 & 0.57 & 3 & 0.86 & 7\\ 
$\gamma$-div, $\gamma=0.05$ & 1.00 & 9 & 0.93 & 8\\ 
$\gamma$-div, $\gamma=0.1$  & 1.00 & 9 & 1.00 & 9\\ 
$\gamma$-div, $\gamma=0.5$  & 1.00 & 9 & 1.00 & 9\\ 
$\beta$-div, $\beta=0.05$ & 0.64 & 4 & 0.71 & 5\\ 
$\beta$-div, $\beta=0.1$  & 0.71 & 5 & 0.79 & 6\\ 
$\beta$-div, $\beta=0.5$  & 1.00 & 9 & NA & NA\\ 
tlasso, $\nu=1$        & 0.86 & 7 & 0.79 & 6\\ 
tlasso, $\nu=5$        & 0.64 & 4 & 0.79 & 6\\ 
nonparanormal          & 0.79 & 6 & 0.79 & 6\\ 
   \hline
\end{tabular*}
\end{table}

\subsection{Gene function regulations}

\begin{figure}[!t]
\centering
\includegraphics[width=16cm]{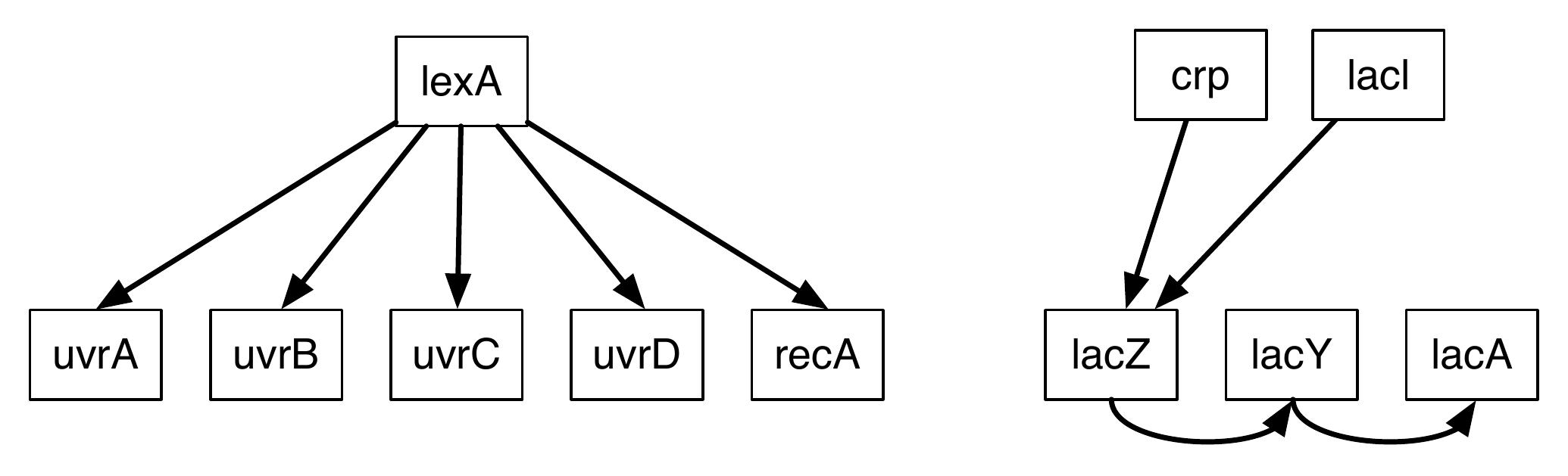}
\caption{Gene function regulations on {\it E.coli} \citep{alberts2014molecular}}.   
\label{fig:yss_truegraph}
\end{figure}

\citet{yamada2014least} selected $p=11$ genes on {\it E.coli} with $n=445$ gene expression levels \citep{faith2007large}, because gene function regulation relationships are well-known, as in Figure \ref{fig:yss_truegraph} \citep{alberts2014molecular}. The most characteristic point is that two network groups exist independently.

\begin{figure}[!t]
\centering
\includegraphics[width=12cm]{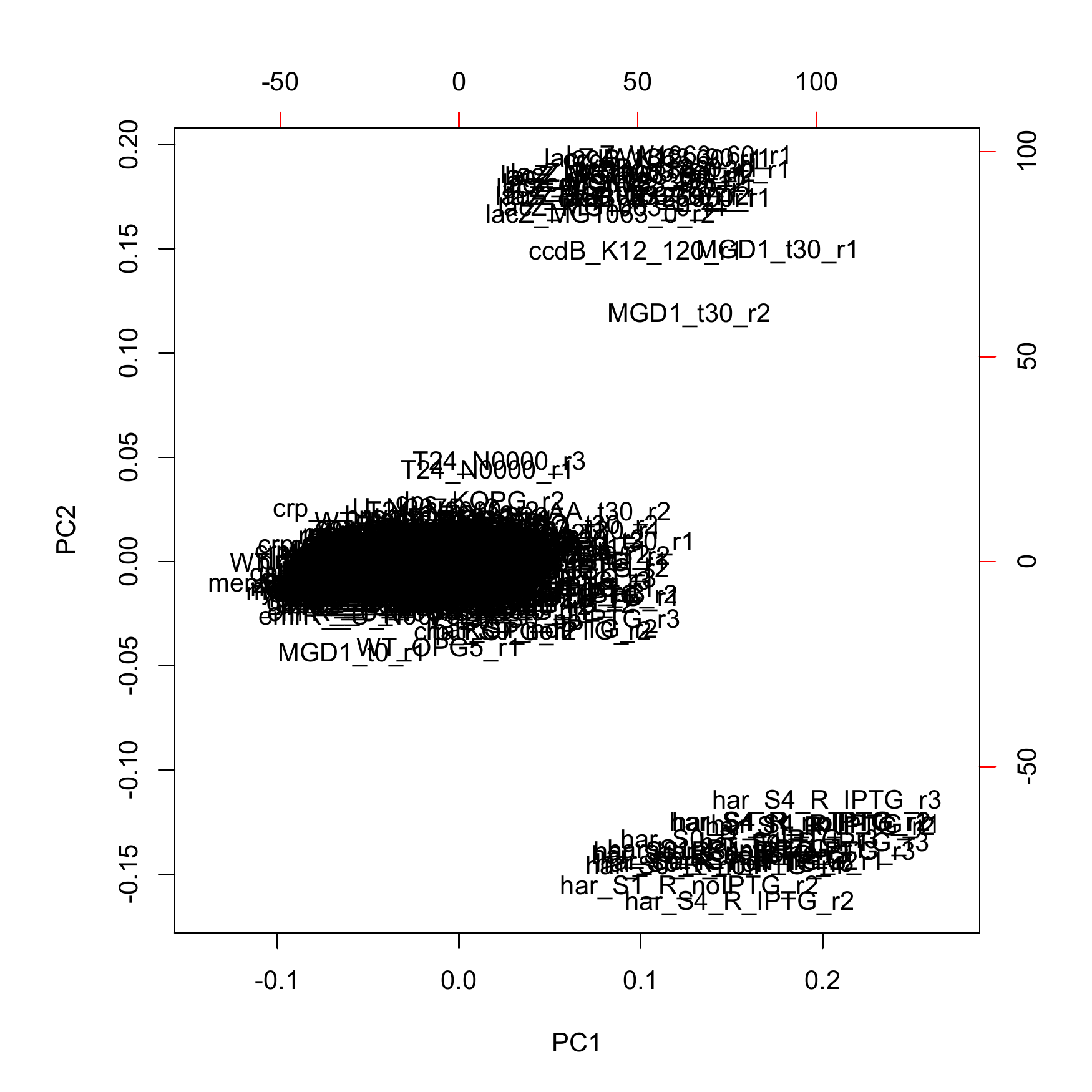} 
\caption{Score plot of PCA.}
\label{fig:biplot_yss}
\end{figure}

We conducted the principal component analysis to the dataset. The data values were normalized in advance with the robust estimates of mean and scale (median and adjusted median absolute deviation (MAD)). The scores of the first two principal components are plotted in Figure \ref{fig:biplot_yss}.  
There exist three clusters in the score plot.  We see that  91.7 \% observations belong to a cluster located in the center of the score plot, and the remaining observations can be regarded as outliers, so that the dataset includes many outliers. 

\begin{figure}[!t]
\centering
\includegraphics[width=16cm]{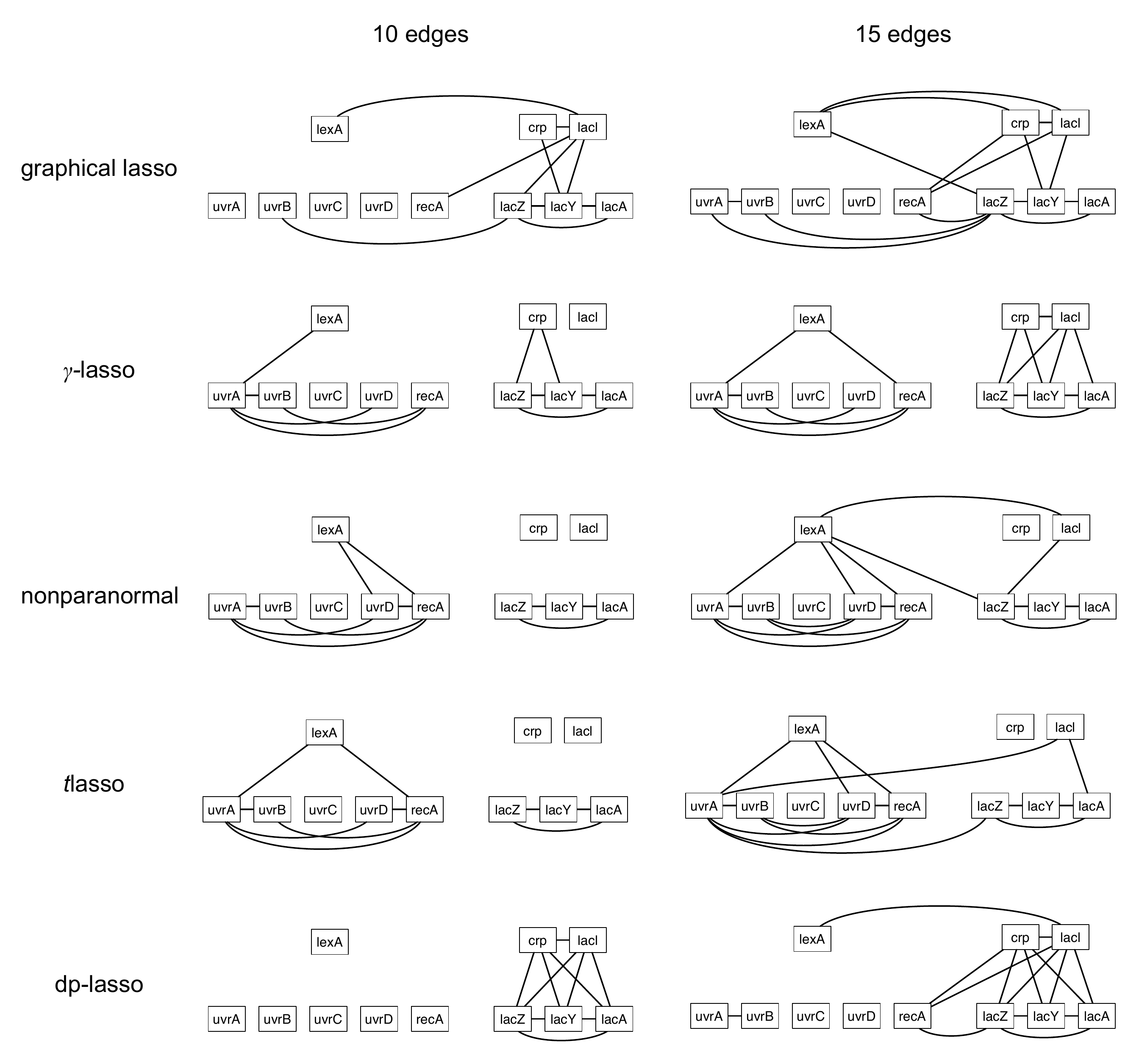}
\caption{Graphical models when the number of edges was 10 and 15.  }
\label{fig:yss_graph}
\end{figure}

Figure \ref{fig:yss_graph} presents graphical models estimated by the graphical lasso, nonparanormal, \tlasso\ with $\nu=1$, the \gmethod\ with $\gamma=0.05$, and the dp-lasso with $\beta=0.05$.    The tuning parameter $\lambda$ was selected so that the number of edges was 10 and 15.  When the number of edges was 10, the graphical lasso produced the edges between two independent groups, but all robust estimation procedures did not. When the number of edges was 15, all methods except for \gmethod\ produced more than one edge connecting between the two groups. Therefore, the \gmethod\ performed the best in terms of estimating fewer cross edges.

\section{Concluding remarks}

We have proposed a robust estimation procedure, \gmethod, based on the $\gamma$-divergence for estimating the high-dimensional graphs.  The parameter estimation procedure was constructed by using the MM algorithm with the standard algorithm used in the graphical lasso.  Extensive simulation studies showed that \gmethod\ performed much better than the existing methods. Two data analyses illustrated that the \gmethod\ worked well. 

The proposed method is based on the $\gamma$-divergence. We can also adopt an extension of the $\gamma$-divergence, H\"{o}lder divergence \citep{KF2014}. The H\"{o}lder divergence allows us to estimate the contamination ratio as well as the model parameters.  A distinguishing point is that we can use an iterative minimization algorithm based on the Pythagorean relation for the H\"{o}lder divergence, which is the same parameter estimation procedure as in the minimization of $\gamma$-divergence \citep{KF2015}. In a similar manner, the proposed parameter estimation procedure based on the MM algorithm is applicable even for the robust sparse Gaussian graphical modeling where the $\gamma$-divergence is replaced by the H\"{o}lder divergence. 

The covariance estimation based on the likelihood function include a wide variety of statistical models such as the factor analysis, the probabilistic principal component analysis \citep{tipping1999probabilistic}, and the canonical correlation analysis.   As a future research topic, it is interesting to extend our method to various covariance estimation procedures.  

Another important topic is the investigation of the asymptotic properties of the \gmethod.  In our simulation study in Section \ref{subsubsec:MSE}, the mean squared error (MSE) of the estimate approached 0 as the number of observations $n$ increased.  In fact, \citet{fujisawa2008robust} showed that the MSE approaches 0 as $n \rightarrow \infty$ even if the contamination ratio is large.   However, the authors showed this property only when the number of variables $p$ is fixed and there is no penalization.  On the other hand, when no outliers are present, the MSE of the graphical lasso is close to zero when both $n$ and $p$ are sufficiently large \citep{rothman2008sparse}.  
Such an asymptotic property in the presence of the outliers is a future issue.    

\section*{Acknowledgement}
The authors would like to thank Dr. Hokeun Sun for providing an {\tt R} code of the dp-lasso.  
\appendix
\def\thesection{Appendix \Alph{section}}
 \section{Derivation of \texorpdfstring{$\ell_2(\bm{\theta})$}{TEXT}}

We calculate $\ell_2(\bm{\theta})$.  $\ell_2(\bm{\theta})$ can be expressed as
\begin{eqnarray}
\ell_2(\bm{\theta}) &=& \frac{1}{1+\gamma} \log \int f(\bm{x};\bm{\theta})^{1+\gamma}d\bm{x}\cr
&=& \frac{1}{1+\gamma} \log \int (2\pi)^{-(1+\gamma)p/2} |\bm{\Omega}|^{(1+\gamma)/2} \exp \left\{ -\frac{1+\gamma}{2}(\bm{x} - \bm{\mu})^T\bm{\Omega}(\bm{x} - \bm{\mu}) \right\} d\bm{x}. \label{解きたい小問題2-2}
\end{eqnarray}
Because the probability of the multivariate normal distribution is one, we have
\begin{eqnarray}
\int  \exp \left\{ -\frac{1+\gamma}{2}(\bm{x} - \bm{\mu})^T\bm{\Omega}(\bm{x} - \bm{\mu}) \right\} d\bm{x} &=& (2\pi)^{p/2} (1+\gamma)^{-p/2} |\bm{\Omega}|^{-1/2}. \label{解きたい小問題2-3}
\end{eqnarray}
Substituting (\ref{解きたい小問題2-3}) into (\ref{解きたい小問題2-2}) gives us
\begin{eqnarray*}
\ell_2(\bm{\theta}) &=&\frac{1}{1+\gamma} \log \left\{  (2\pi)^{-(1+\gamma)p/2} |\bm{\Omega}|^{(1+\gamma)/2} (2\pi)^{p/2} (1+\gamma)^{-p/2} |\bm{\Omega}|^{-1/2} \right\}\\
&=& \frac{1}{1+\gamma} \log \left\{  (2\pi)^{-\gamma p/2} |\bm{\Omega}|^{\gamma/2} (1+\gamma)^{-p/2} \right\}\\
&=& \frac{1}{1+\gamma} \left\{ - \frac{\gamma p}{2}\log  (2\pi) +  \frac{\gamma}{2}\log|\bm{\Omega}| -\frac{p}{2} \log(1+\gamma)\right\}. 
\end{eqnarray*}

\section{Determination of $\lambda_1$}
The value of $\lambda_1$, which is the minimum value of $\lambda$ so that all of the non-diagonal elements of inverse covariance matrix are zeros, is easily obtained.  When $\omega_{jk} = 0$ for any $j \ne k$, $\sigma_{jk} = 0$ as well.  Therefore, the variance for each variable is estimated separately.  For $j$th variable, the mean $\mu_j$ and variance $\sigma_{jj}$ are estimated by the iterative algorithm based on Example 4.1 of \citet{fujisawa2008robust}:
\begin{eqnarray*}
	\mu_j^{(t+1)} = \sum_{i=1}^nw_i^{(t)}x_{ij}, \quad \sigma_{jj}^{(t+1)} = \frac{1}{1+\gamma} \sum_{i=1}^n w_i^{(t)} (x_{ij} - \mu_j^{(t+1)})^2,
\end{eqnarray*}
where $w_i^{(t)}$ is the weight given by (\ref{weight_gamma}). Let $\hat{\bm{\mu}}$ and $\hat{\sigma}_{jj}$ be the estimate of $\bm{\mu}$ and $\sigma_{jj}$ obtained by the above algorithm, respectively.  Let the weight in (\ref{weight_gamma}) based on $\hat{\bm{\mu}}$ and $\hat{\sigma}_{jj}$ be denoted by $\hat{w}_i$.  We obtain the following weighted sample covariance matrix:
\begin{eqnarray*}
\bm{S}_{w} = \sum_{i=1}^n \hat{w}_i(\bm{x}_i - \hat{\bm{\mu}})(\bm{x}_i - \hat{\bm{\mu}})^T. 
\end{eqnarray*}

The value of $\lambda_1$ is then estimated by $\lambda_{1} = \|\bm{S}_w - {\rm Diag}(\bm{S}_w) \|_{\infty}$. The basic idea is a necessary and sufficient condition for the solution to the graphical lasso problem given by Corollary 1 of \citet{witten2011new}.

\bibliographystyle{ECA_jasa} 
\bibliography{paper-ref}
\end{document}